\documentclass{emulateapj}	% Nice 2-col format for preprints

\newcommand{\msun}{\mbox{${\cal M}_{\odot}$}}	% Solar mass
\newcommand{\msunyr}{\mbox{${\cal M}_{\odot}/{\rm yr}$}}	% Solar mass
\newcommand{\n}{NGC~}				% NGC = New General Catalogue
\newcommand{\zo}{Z{$_\odot$}}			% Solar metallicity
\newcommand{\rprime}{\mbox{$r^\prime$}}         % r' magnitude
\newcommand{\gprime}{\mbox{$g^\prime$}}         % g' magnitude
\shorttitle{NGC~4038/39  Clusters}
\shortauthors{Bastian et al.}

\begin{document}

\title{Gemini Spectroscopic Survey of Young  Star Clusters in Merging/Interacting Galaxies. III. The Antennae}  

%% Use \author, \affil, and the \and command to format
%% author and affiliation information.
%% Note that \email has replaced the old \authoremail command
%% from AASTeX v4.0. You can use \email to mark an email address
%% anywhere in the paper, not just in the front matter.
%% As in the title, use \\ to force line breaks.

\author{Nate Bastian}
\affil{Institute of Astronomy, University of Cambridge, Madingley Road, Cambridge, CB3 0HA, United Kingdom}
\email{bastian@ast.cam.ac.uk}
\author{Gelys Trancho}
\affil{Gemini Observatory, Casilla 603, La Serena, Chile}
%\affil{Gemini Observatory, 670 N. A'ohoku Place, Hilo, HI 96720, USA}
\email{gtrancho@gemini.edu}
%\and
\author{Iraklis S. Konstantopoulos}
\affil{Department of Physics and Astronomy, University College London, Gower Street, London, WC1E 6BT, UK}
\affil{Gemini Observatory, Casilla 603, La Serena, Chile}

\and
\author{Bryan W. Miller}
\affil{Gemini Observatory, Casilla 603, La Serena, Chile}

%% 
%% Mark off your abstract in the ``abstract'' environment. In the manuscript
%% style, abstract will output a Received/Accepted line after the
%% title and affiliation information. No date will appear since the author
%% does not have this information. The dates will be filled in by the
%% editorial office after submission.

\begin{abstract}

We present optical spectroscopy of 16 star clusters in the merging galaxies NGC~4038/39 ("The Antennae") and supplement this dataset with %U, B, V, I, \& H$\alpha$ 
HST imaging.  The age and metallicity of each cluster is derived through a comparison between the observed Balmer and metal line strengths with simple stellar population models.  We then estimate extinctions and masses using the photometry.  We find that all but three clusters have ages between $\sim3-200$~Myr, consistent with the expected increase in the star-formation rate due to the merger.    Most of the clusters have velocities in agreement with nearby molecular and H{\sc i} gas that has been previously shown to be rotating within the progenitor galaxies, hence star/cluster formation is still taking place within the galactic disks.  However, three clusters have radial velocities that are inconsistent with being part of the rotating gas disks, which is surprising given their young ($200-500$~Myr) ages.  Interestingly, we find a stellar association with the same colors (V-I) near one of these three clusters, suggesting that the cluster and association were formed concurrently and have remained spatially correlated.  We find evidence for spatially distributed cluster formation throughout the duration of the merger.  The impact of various assumptions about the star/cluster formation rate on the interpretation of the cluster age distribution are explored, and we do not find evidence for long term "infant mortality" as has been previously suggested.  Models of galaxy mergers that include a prescription for star formation  can provide an overall good fit to the observed cluster age distribution.

\end{abstract}

%% Keywords should appear after the \end{abstract} command. The uncommented
%% example has been keyed in ApJ style. See the instructions to authors
%% for the journal to which you are submitting your paper to determine
%% what keyword punctuation is appropriate.

%% Authors who wish to have the most important objects in their paper
%% linked in the electronic edition to a data center may do so in the
%% subject header.  Objects should be in the appropriate "individual"
%% headers (e.g. quasars: individual, stars: individual, etc.) with the
%% additional provision that the total number of headers, including each
%% individual object, not exceed six.  The \objectname{} macro, and its
%% alias \object{}, is used to mark each object.  The macro takes the object
%% name as its primary argument.  This name will appear in the paper
%% and serve as the link's anchor in the electronic edition if the name
%% is recognized by the data centers.  The macro also takes an optional
%% argument in parentheses in cases where the data center identification
%% differs from what is to be printed in the paper.


%% From the front matter, we move on to the body of the paper.
%% In the first two sections, notice the use of the natbib \citep
%% and \citet commands to identify citations.  The citations are
%% tied to the reference list via symbolic KEYs. The KEY corresponds
%% to the KEY in the \bibitem in the reference list below. We have
%% chosen the first three characters of the first author's name plus
%% the last two numeral of the year of publication as our KEY for
%% each reference.

\section{INTRODUCTION}
\label{sec:intro}

The formation of large numbers of massive star clusters in merging and starburst galaxies is now a commonly observed phenomenon.  The clusters that form in such mergers retain information of the history of the galaxy.  This fact is often exploited to investigate how and when a galaxy formed, specifically to test if (and what fraction) of elliptical galaxies formed via the merger of gas rich spirals (e.g. Schweizer~1987; Ashman \& ~Zepf~1992).

A significant amount of work has been done on cluster populations of intermediate aged merger remnants (NGC 7252 - Miller et al.~1997, Schweizer \& Seitzer~1998; NGC 1316 - Goudfrooij et al.~2001; NGC 1700/3610 - Whitmore et al.~1997 ), with the main goal of using the cluster populations to trace the merger and subsequent evolution.  While older mergers are generally easier to interpret, as much of the star-formation has ceased and the merger remnant has become less dust obscured, it is the earlier stages of the merger where a lot of the key aspects of the cluster population are set.  What are the initial properties of the population?  How many clusters are likely to survive to old ages that can be seen in the aging remnants?  It is in this phase where the study of cluster populations in their own right and their use as galaxy evolution tracers begin to overlap.

We have been conducting a spectroscopic survey of star clusters in a series of young/on-going galactic mergers and interactions to address both the galactic transformation and the formation of the cluster population.  Our first target was NGC~3256 (Trancho et al.~2007a,b; hereafter T07a,b), a relatively distant pair of gas-rich spirals ($\sim36$~Mpc) that are in the advanced stages of merging, although the two nuclei are still separate.  The kinematics of the clusters showed that they were still clearly associated with a molecular gas disk, which has not yet been destroyed by the merger.    A somewhat surprising find presented in T07a was the discovery of three relatively old ($60-200$~Myr) clusters in the tidal debris of the merger.  The ages and velocities of the clusters led us to conclude that they must have formed in the tidal tails.

Arguably, the most heavily studied young ongoing merger is the nearby galaxy pair NGC~4038/39 (The Antennae).   Numerical modeling of the merger suggests that the first close passage of the two progenitor disk galaxies occurred $\sim200-400$~Myr ago (e.g. Barnes~1988; Mihos et al.~1993). Early WFPC HST imaging revealed a large population of young clusters (Whitmore \& Schweizer~1995) whose properties have been the subject of many subsequent studies (e.g. Whitmore et al.~1999, Mengel et al.~2005, Anders et al.~2007).  The currently forming clusters have masses up to a few times $10^6$~\msun, and are distributed throughout the galaxy (Zhang et al.~2001).  Due to the vast population of clusters compared with most other galaxies whose populations have been studied, the Antennae clusters have long been used as a test case for the process of cluster formation and dissolution.  

 Fall et al.~(2005; hereafter FCW05) and Whitmore et al.~(2007, hereafter WCF07) have used the measured age distribution to suggest that the vast majority of clusters that are formed do not live to old ages.  They suggest that 90\% of clusters are disrupted every dex of age, a process that continues between a few Myr to a $\sim100$~Myr.  This leads to an age distribution where the number of clusters in a mass limited sample decreases as $\tau^{-1}$.  %FCW05 show that the age distribution of clusters in the Antennae follows the $\tau^{-1}$ for approximately 1~Gyr (their Fig.~2).  
This decrease has been interpreted as being due to "infant mortality", thought to be caused  by the removal of natal gas from the cluster.  Modeling of this process suggests that its effects should be largely over by $10-20$~Myr (Goodwin \& Bastian~2006; Baumgardt \& Kroupa~2007)\footnote{In fact, $10-20$~Myr is an upper limit, as it depends on the crossing time of the clusters, which in turn depends on the core radius.  In Goodwin \& Bastian~(2006) an initial core radius of 1~pc was used, which is much larger than the core radii observed for young ($<10$~Myr) massive clusters (Bastian et al.~2008).}, although we note that these models do not take stellar evolution into account.  For clarity, we will refer to the FCW05 and WCF07 disruption model as "long duration infant mortality" to distinguish it from the original term "infant mortality" (Lada \& Lada~2003) meant to describe the disruption of clusters as they pass from an embedded to an exposed phase.

One possible cause of the difference between the long duration infant mortality model and simulations, which we explore in this work, is that FCW05 have assumed that the cluster formation rate within this merging galaxy has been constant for the past Gyr.  This will severely affect the interpretation of the data as the observed age distribution is a convolution of the formation history and cluster disruption.

There has been some debate about the distance to the Antennae, with the recession velocity leading to a distance of $\sim20$~Mpc (assuming H$_{0}$=72 km/s/Mpc) and suggestions that the tip of the red giant branch lead to a distance of $13.3$~Mpc (Saviane et al.~2008).  Schweizer et al.~(2008) have used the type-Ia supernova 2007sr to estimate a distance of 22.3~Mpc.  In the present work, we adopt a distance of 20~Mpc, in accord with the supernova data, but also to be able to compare directly with previous works.

This paper is organized in the following way.  In \S~\ref{sec:obs} we introduce the spectroscopic and photometric datasets used and we outline our analysis procedure in \S~\ref{sec:properties}.   Our results are presented in \S~\ref{sec:results} and in \S~\ref{sec:discussion} we interpret the results in light of other studies of cluster populations in merging galaxies.  We summarize our main conclusions in \S~\ref{sec:conclusions}.

\section{OBSERVATIONS AND REDUCTION}\label{sec:obs}

\subsection{Gemini-GMOS Spectroscopy}
\label{sec:spectroscopy}

We  obtained spectra of 20 sources in NGC~4038/39 using the Multi-Object Spectroscopy (MOS) mode of the Gemini Multi-Object Spectrograph (GMOS) on Gemini North.  The entire central region of the merger fits within the 5.5 by 5.5  arcminute GMOS field-of-view.   Imaging of the galaxy and spectroscopy of the selected star clusters were obtained with GMOS-N in semester 2003A. The data were obtained as part of Queue program GN-2003A-Q-33.  Pre-imaging was obtained through two filters, \gprime and \rprime.  The selection of star cluster candidates was based on the derived color magnitude diagram from the pre-imaging, and supplemented by HST imaging in order to confirm their status as clusters.  Two GMOS masks were used for the spectroscopy. 

We used the B600 grating and a slit width of 0.75~arcsec, resulting in a instrumental resolution of 110 km/s at 5000 \AA. The spectroscopic observations were obtained as 8 individual exposures with an exposure time of 1800 sec each.    No atmospheric dispersion corrector (ADC) was available on Gemini-North at the time of the observations, causing wavelength-dependent slit losses.  We corrected each exposure for this effect using the method of Filippenko~(1982).

\begin{figure*}
 \begin{center}
   \epsscale{0.9}
 \plotone{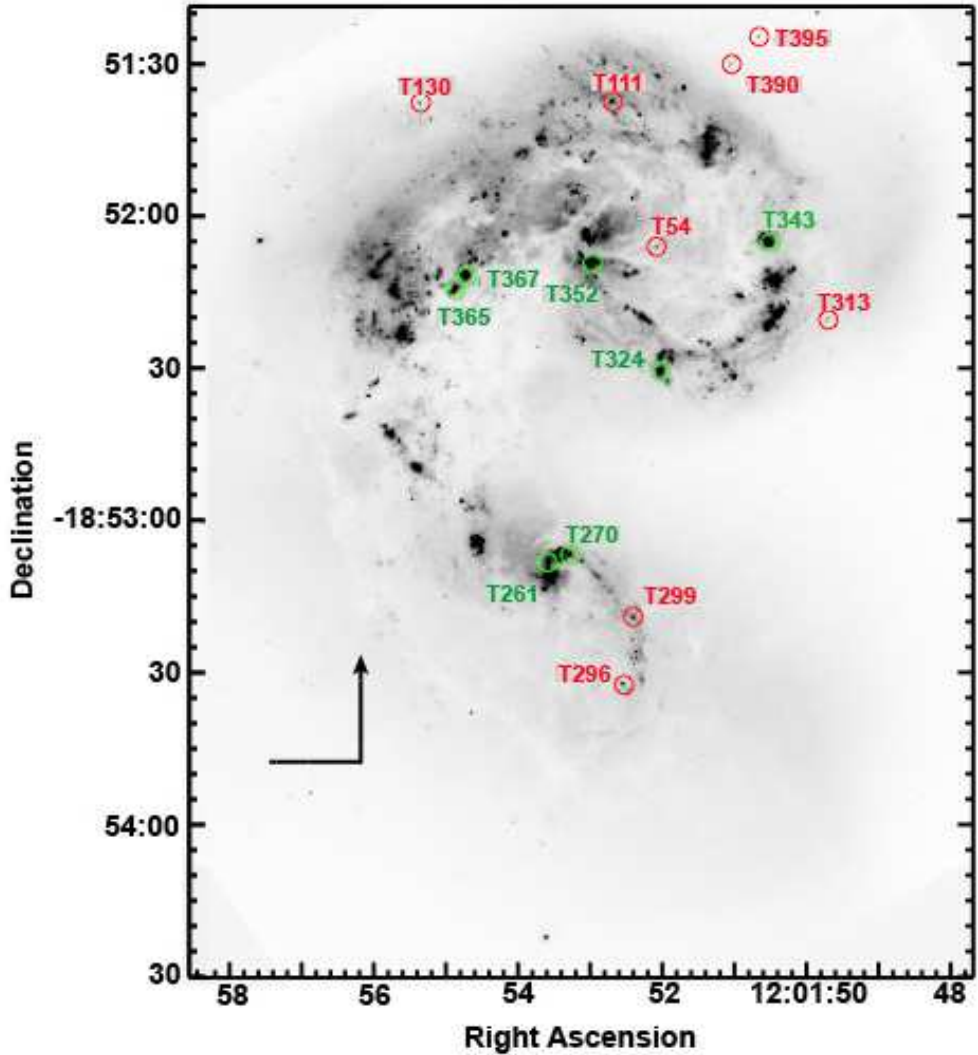}
     \caption{HST ACS B-band image of the main body of the Antennae, showing the locations of the observed clusters. Red represents clusters that display an absorption dominated spectrum, while green represents emission dominated spectra.  A close up image of each cluster is shown in Fig.~\ref{fig:images}.} 
         \label{fig:image}
      \end{center} 

 \end{figure*}

\begin{figure*}
 \begin{center}
   \epsscale{0.9}
 \plotone{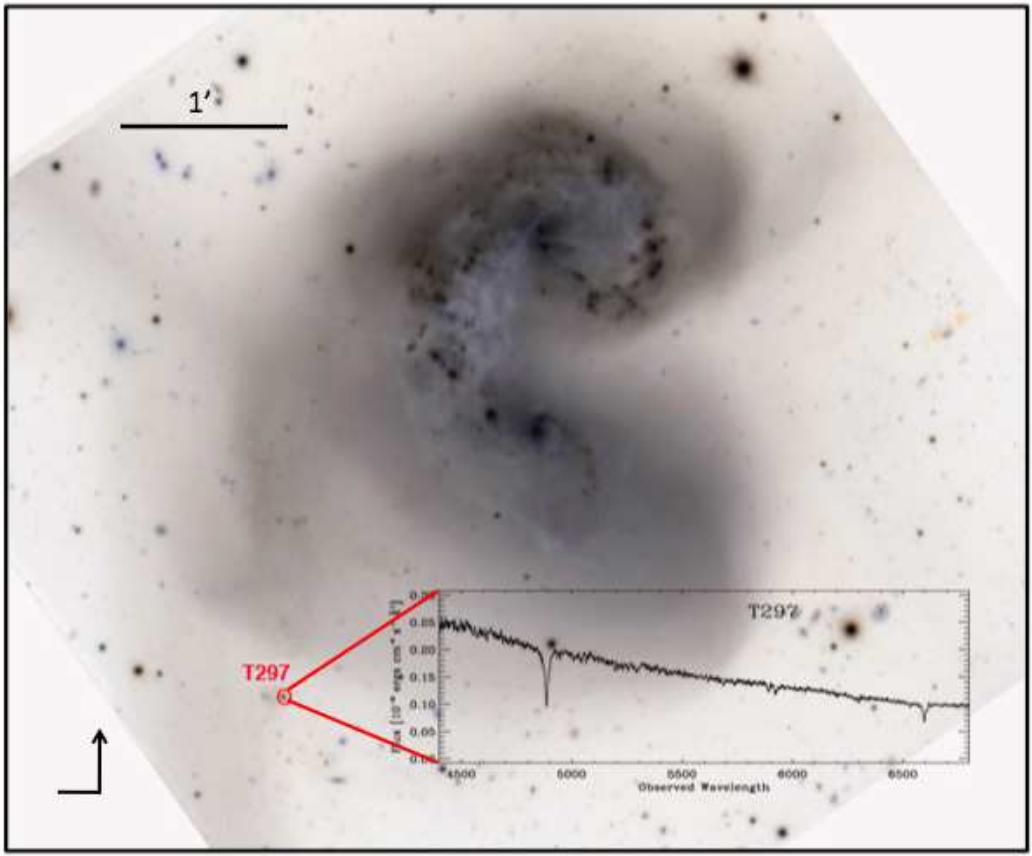}
     \caption{Gemini \gprime\ and \rprime\ image of the Antennae marking the position of the discovered cluster, T297, projected onto the southern tidal tail.  A close up image of this cluster and its adjacent stellar association is shown in Fig.~\ref{fig:images}.  The image has the same orientation as Fig.~\ref{fig:image}.} 
         \label{fig:image-tails}
      \end{center} 
 \end{figure*}

Spectroscopy of 20 candidates yielded 16 that were star clusters with one in the tidal debris.  An additional young cluster in the main body of the galaxy had heavily saturated emission lines, making the extracted spectrum unusable.   Of the remaining sources, two were foreground stars and one was a background quasar.  In this paper we will focus on these star clusters, the IDs and positions are given in Table~\ref{table:colors}.  Figure~\ref{fig:image} shows an {\em HST}/ACS image of the main body of \n4038, with the observed candidate clusters (except for cluster T297, located in the tidal debris) marked by their ID numbers. The position of T297 is given in Fig.~\ref{fig:image-tails}.

\begin{figure*}
 \begin{center}
   \epsscale{0.9}
 \plotone{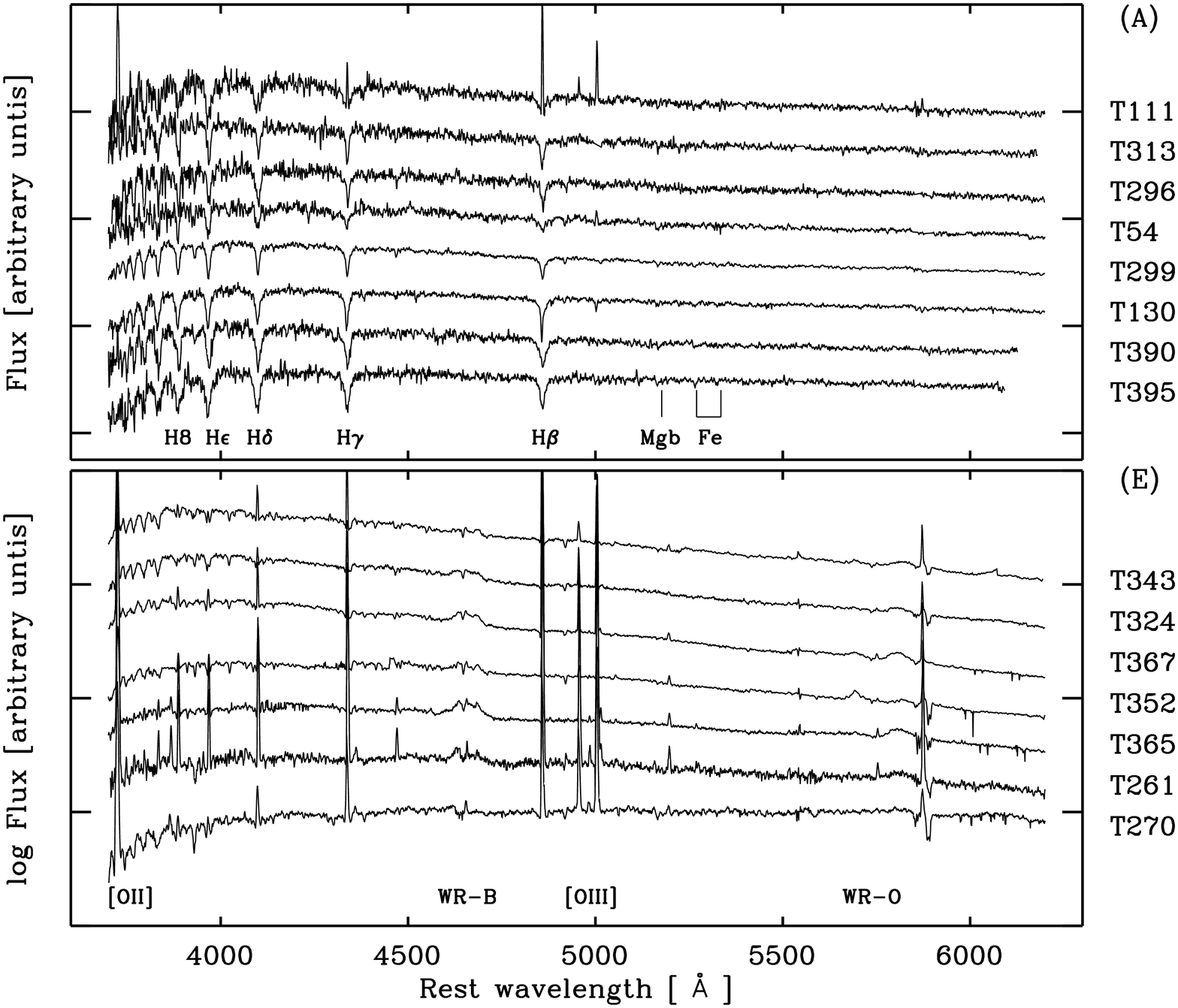}
     \caption{Doppler corrected spectra of clusters presented in this work.  The top and bottom panels show absorption (A) and emission (E) line dominated spectra, respectively.   We have indicated prominent emission and absorption lines used in the analysis.  Wolf-Rayet features are shown as "WR-B" (the $\lambda4650$\AA\  blue bump) and "WR-O" (the $\lambda5800$\AA\ orange bump). Spectra of T390, T111, and T261 are shown in detail in Fig.~\ref{fig:examples}.  Additionally, the spectrum of T297 (the cluster in the tidal debris discussed in \S~\ref{sec:297}) is shown in Fig.~\ref{fig:image-tails}. 
     } 
         \label{fig:spectra}
      \end{center} 
 \end{figure*}

The basic reductions of the data were done using a combination of the Gemini IRAF package and custom reduction techniques, as described in Appendix A in T07b.  The resulting spectra are shown in Fig.~\ref{fig:spectra} where we have separated them into absorption dominated (top panel) and emission dominated (bottom panel).  Additionally, the spectrum of T297 is shown in the inset of Fig.~\ref{fig:image-tails}.

\subsection{HST photometry}
\label{sec:photometry}

Hubble Space Telescope (HST) Advanced Camera for Surveys (ACS) F435W, F550M, F658N, and F814W images of the Antennae were taken from the Hubble Legacy archive (prop ID 10188, PI Whitmore).  The images came fully reduced and drizzled by the standard automatic pipeline. Additionally, we used F336W images from WFPC2, described in Bastian et al.~(2006a).   Photometry was performed on the images with apertures of the same size as the slit width in order to directly compare the spectroscopic and photometric results.  The aperture and inner/outer background radii for the photometry was 7.5/9.5/11.5 pixels in the ACS images (and WFPC2 PC image), respectively.  For the WFCP2 wide field chip images we used 4/5/6 pixels for the aperture/inner/outer-background.  In the majority of cases, there was only a single cluster (or a single dominant cluster) in the extraction aperture.  

\begin{figure*}
 \begin{center}
   \epsscale{0.9}
\includegraphics[width=4cm]{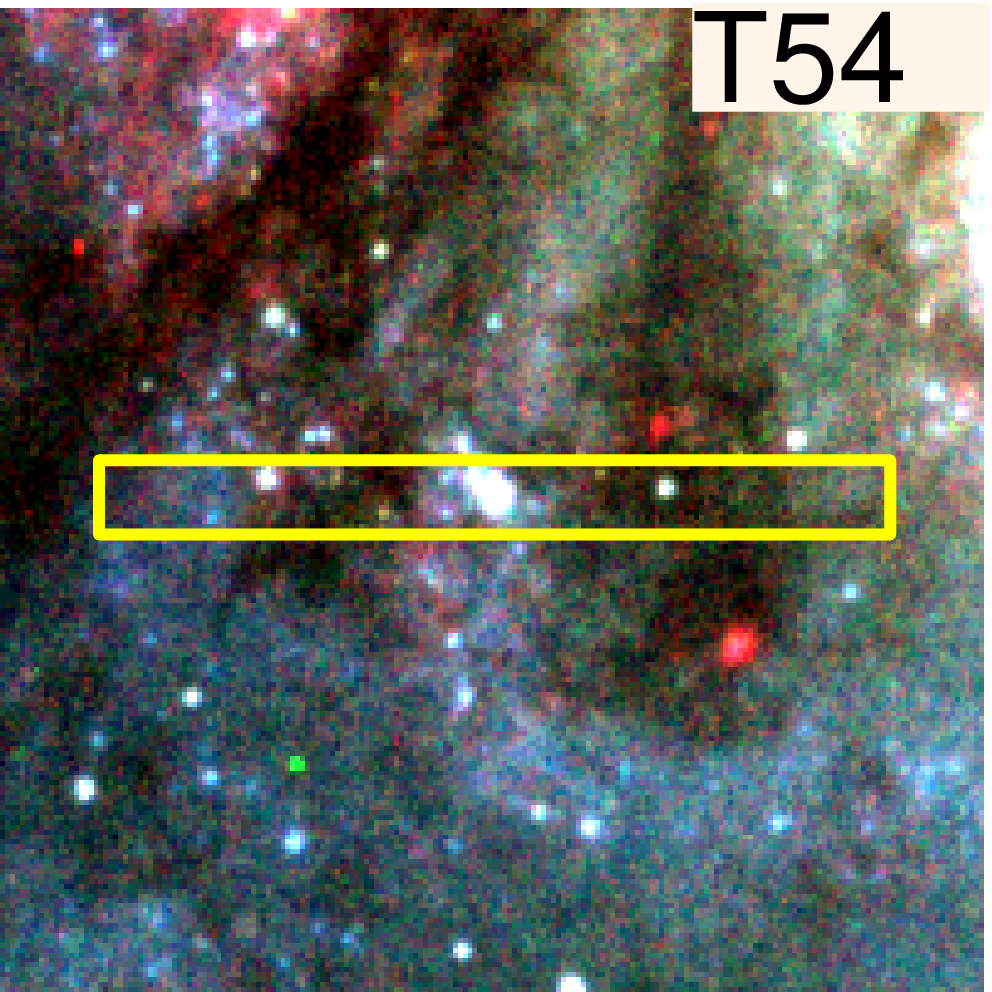}
\includegraphics[width=4cm]{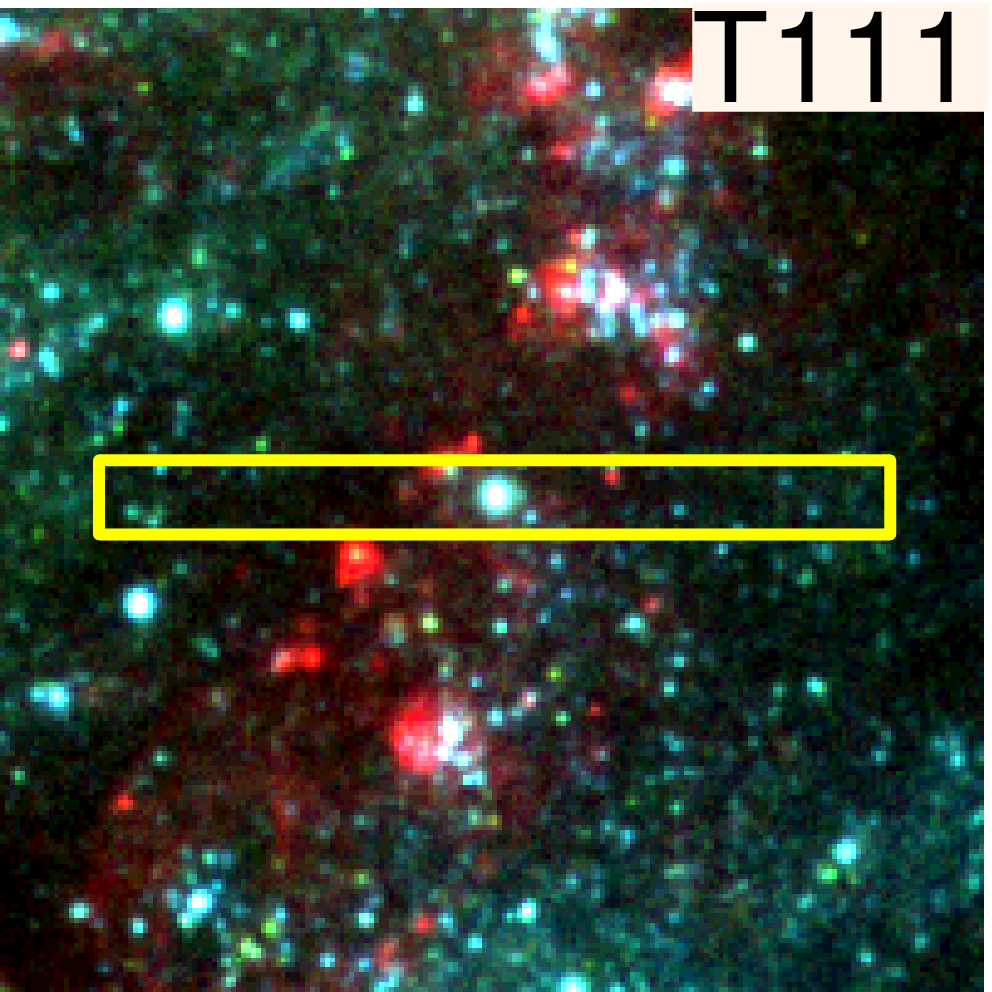}
\includegraphics[width=4cm]{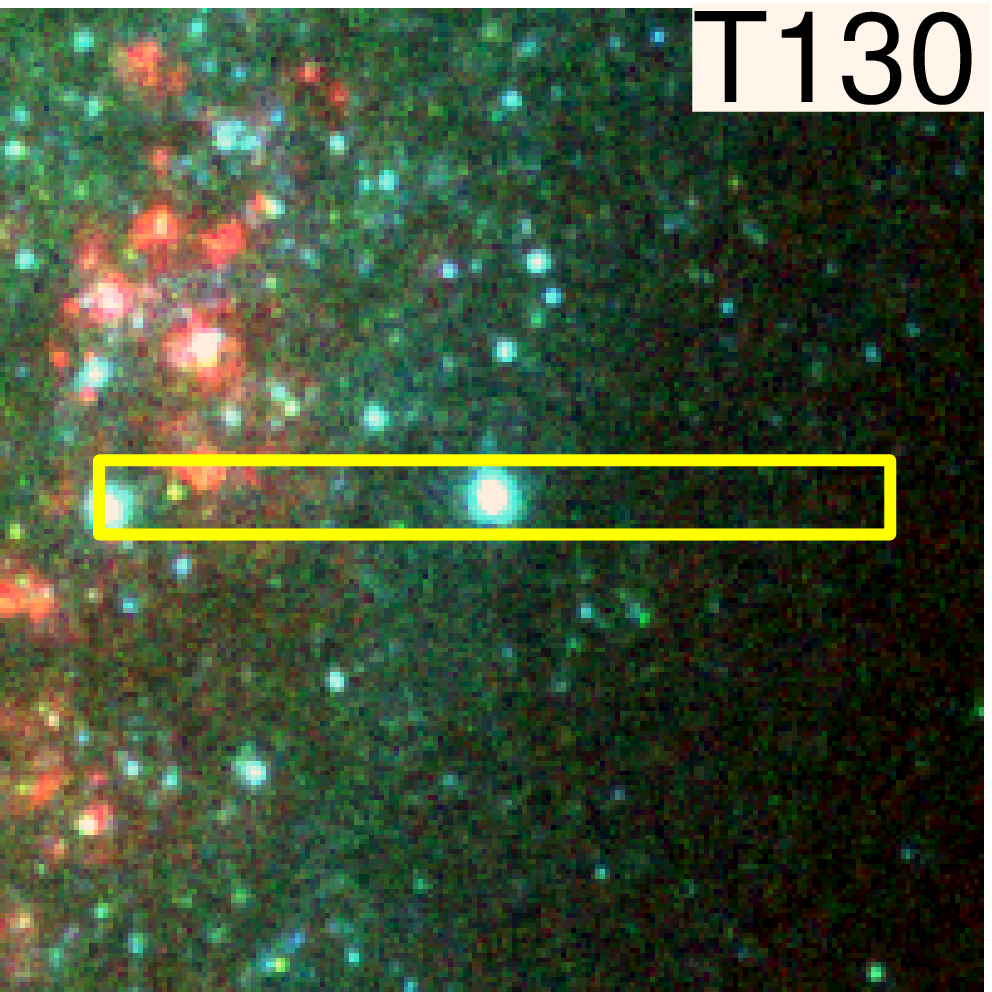}
\includegraphics[width=4cm]{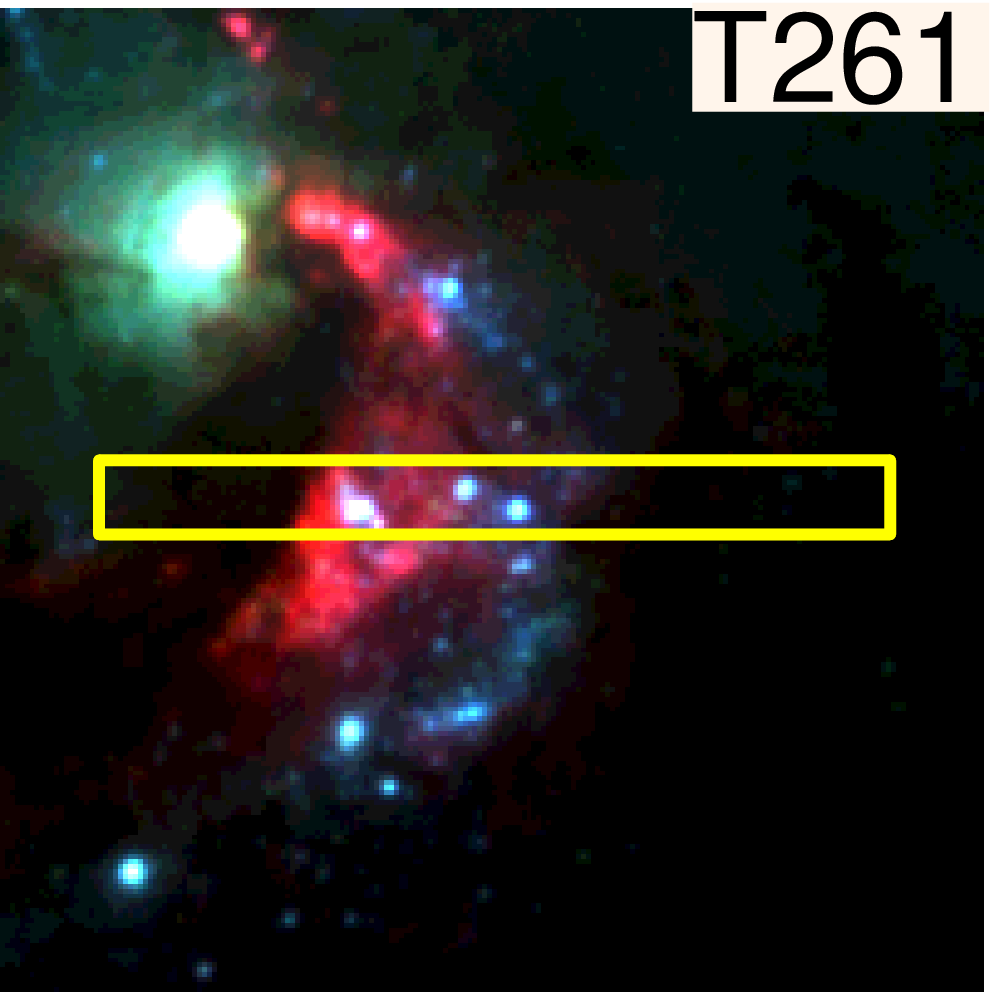}
\includegraphics[width=4cm]{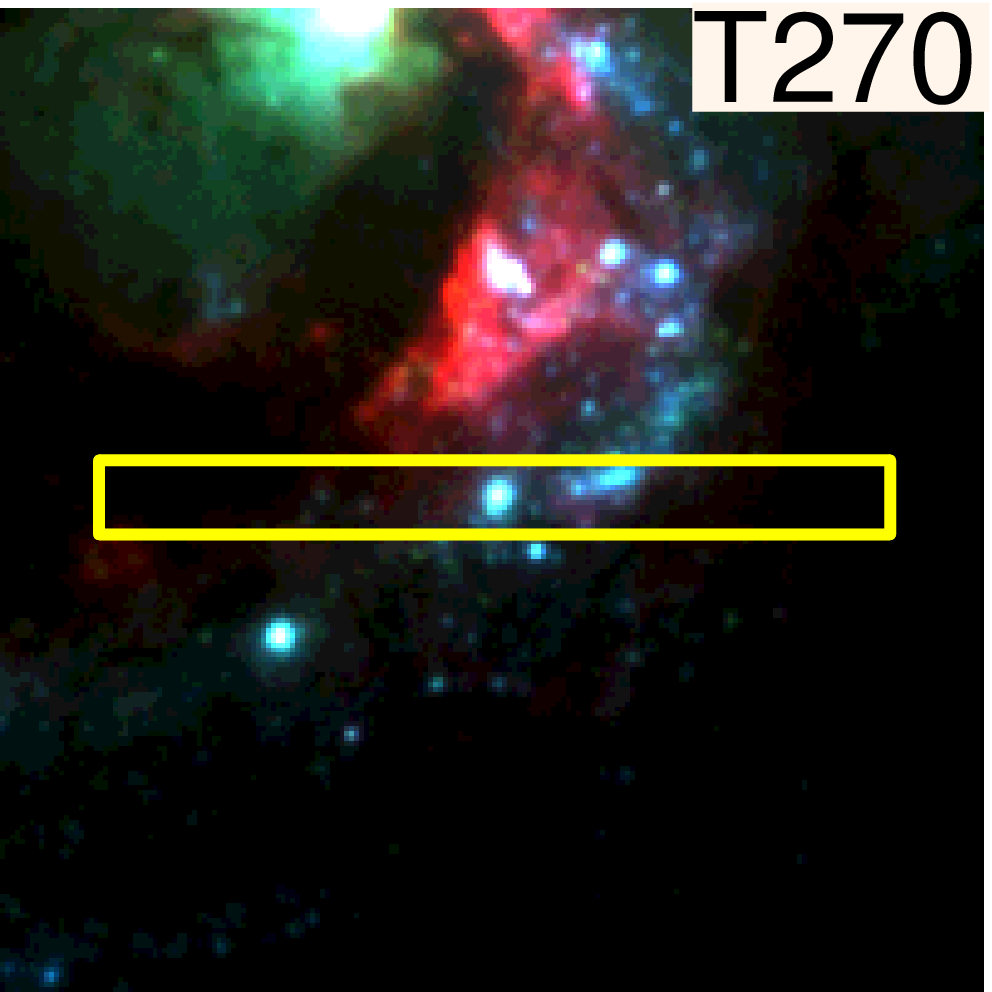}
\includegraphics[width=4cm]{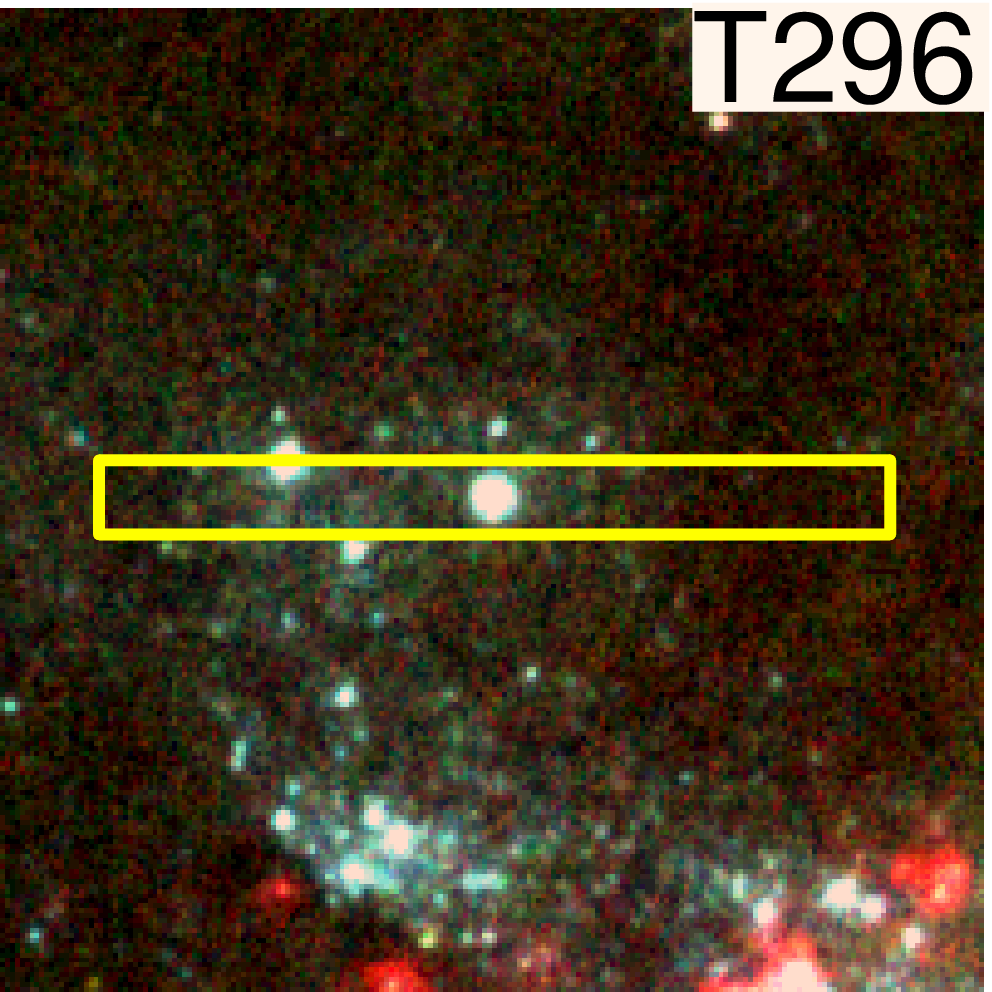}
\includegraphics[width=4cm]{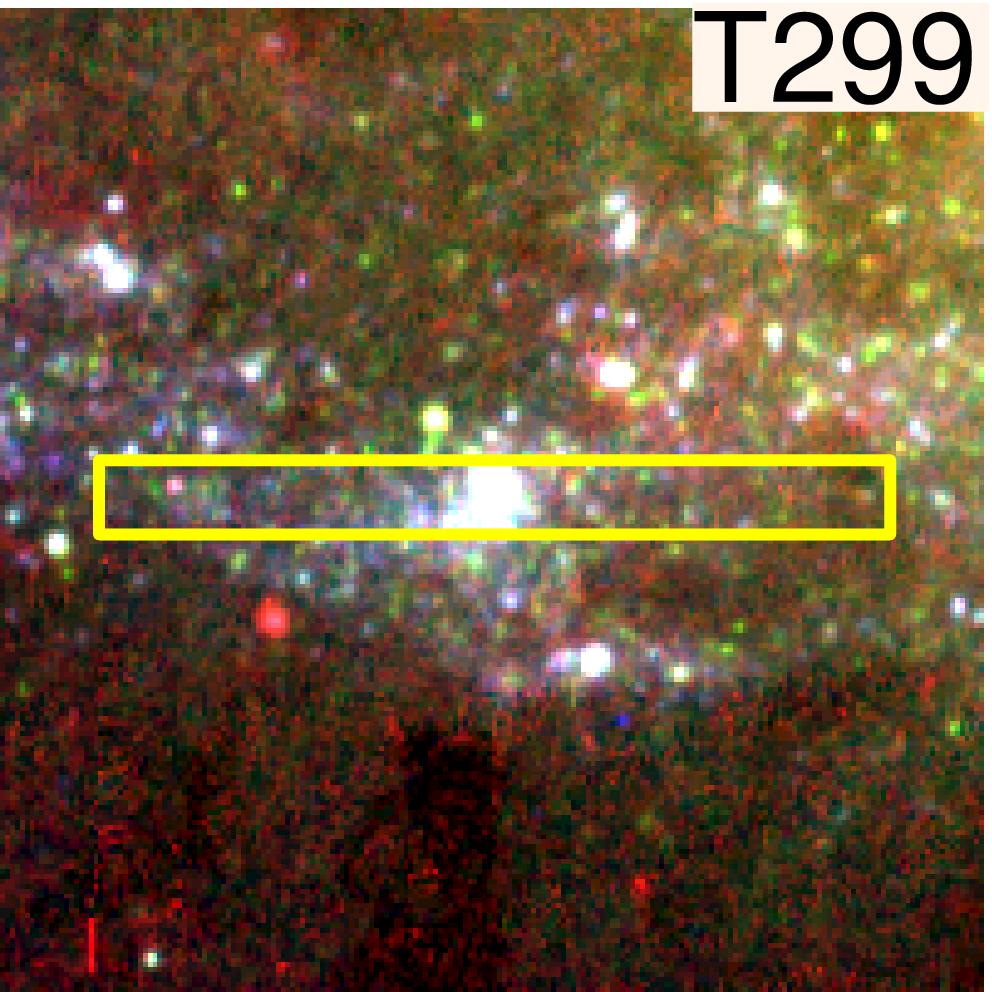}
\includegraphics[width=4cm]{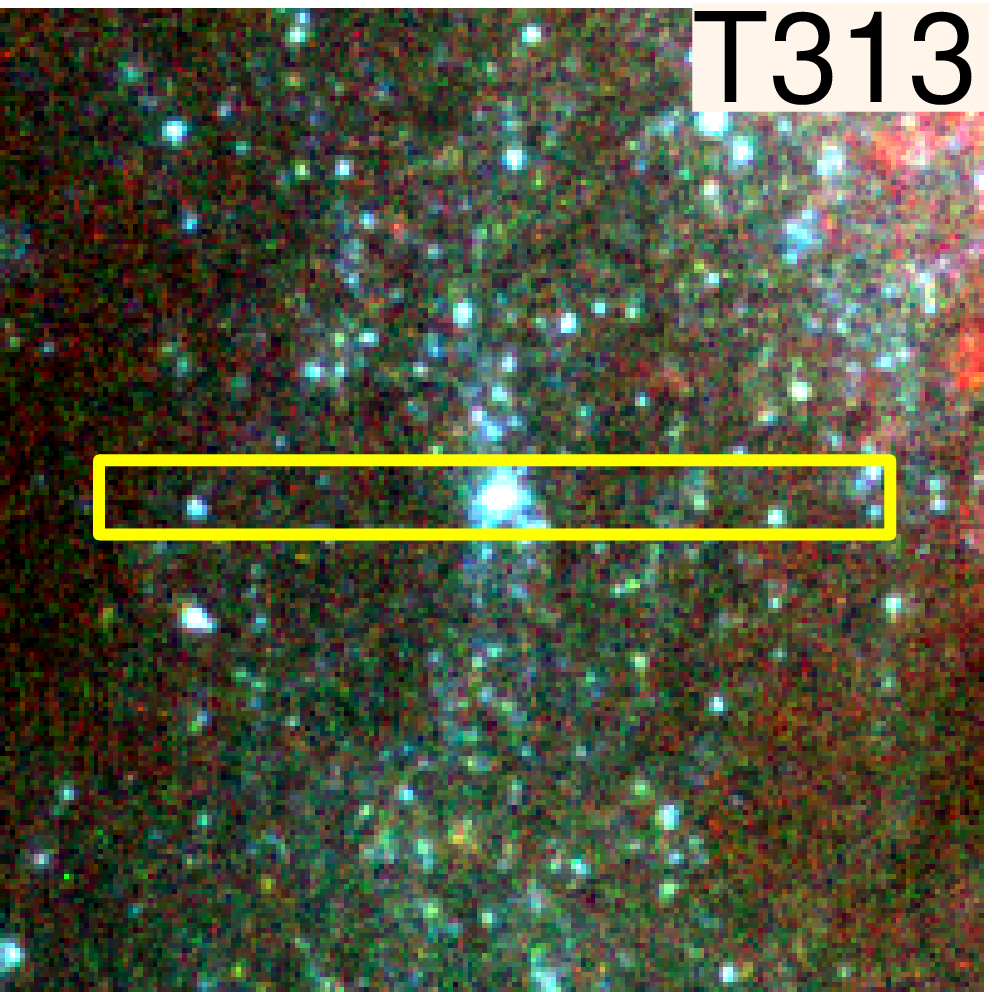}
\includegraphics[width=4cm]{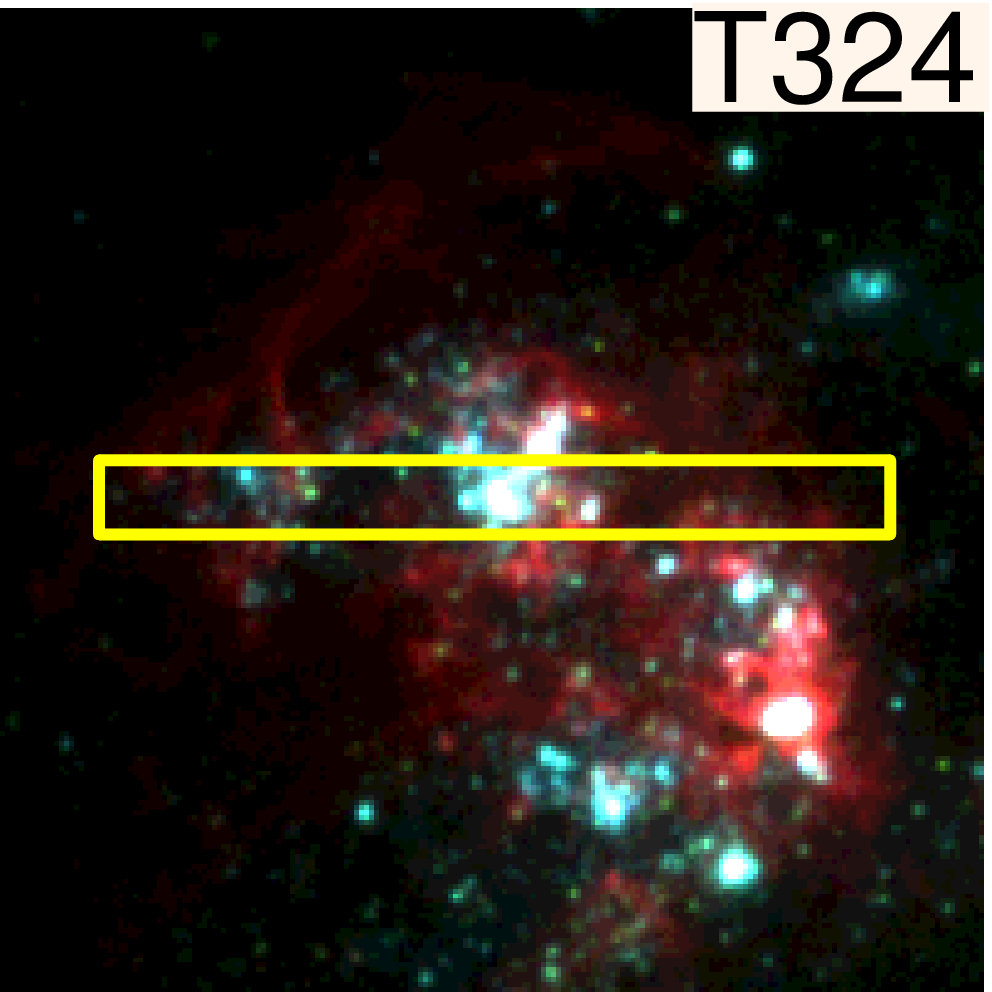}
\includegraphics[width=4cm]{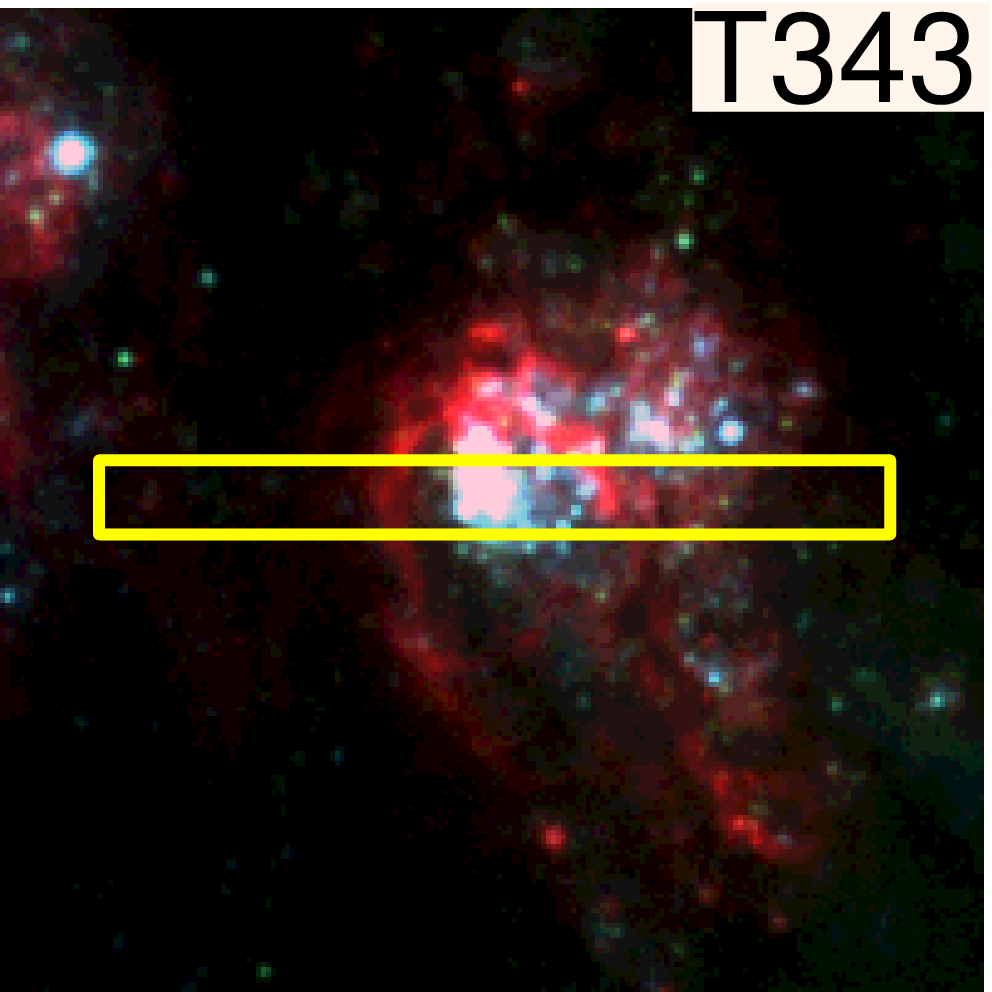}
\includegraphics[width=4cm]{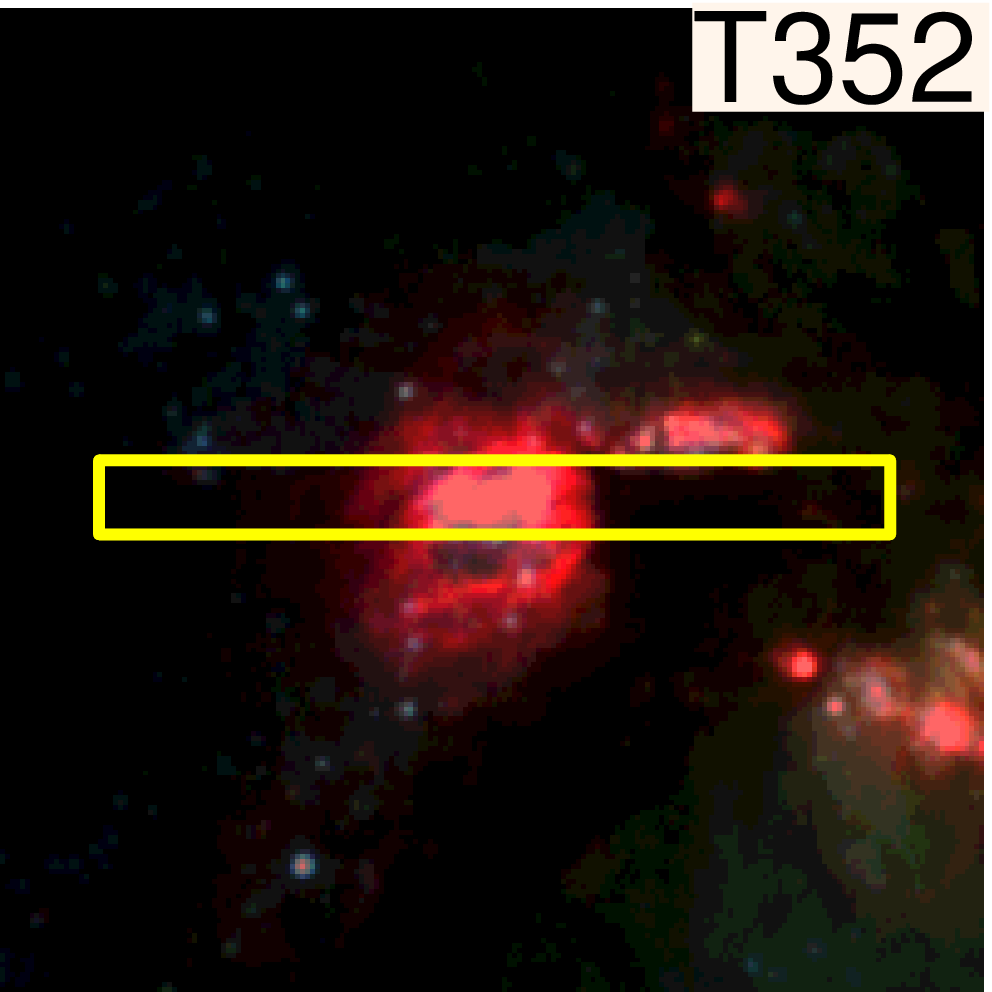}
\includegraphics[width=4cm]{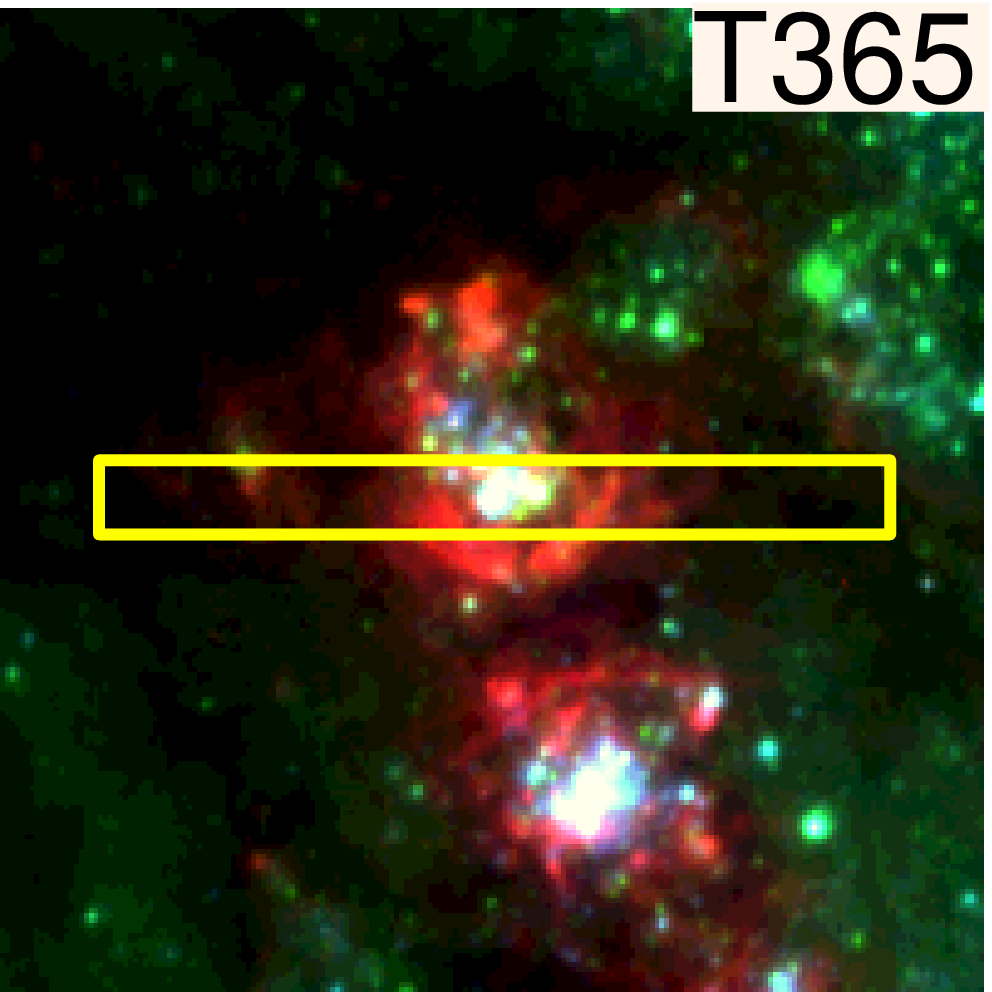}
\includegraphics[width=4cm]{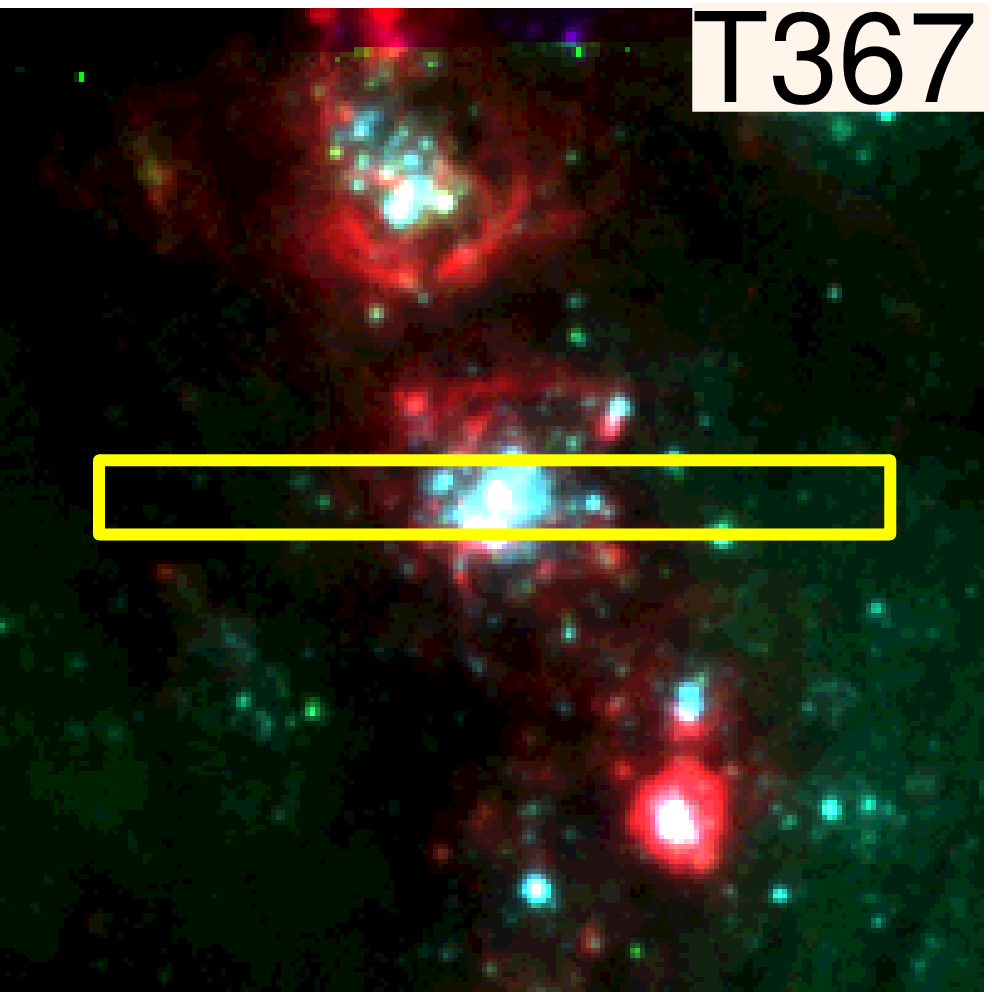}
\includegraphics[width=4cm]{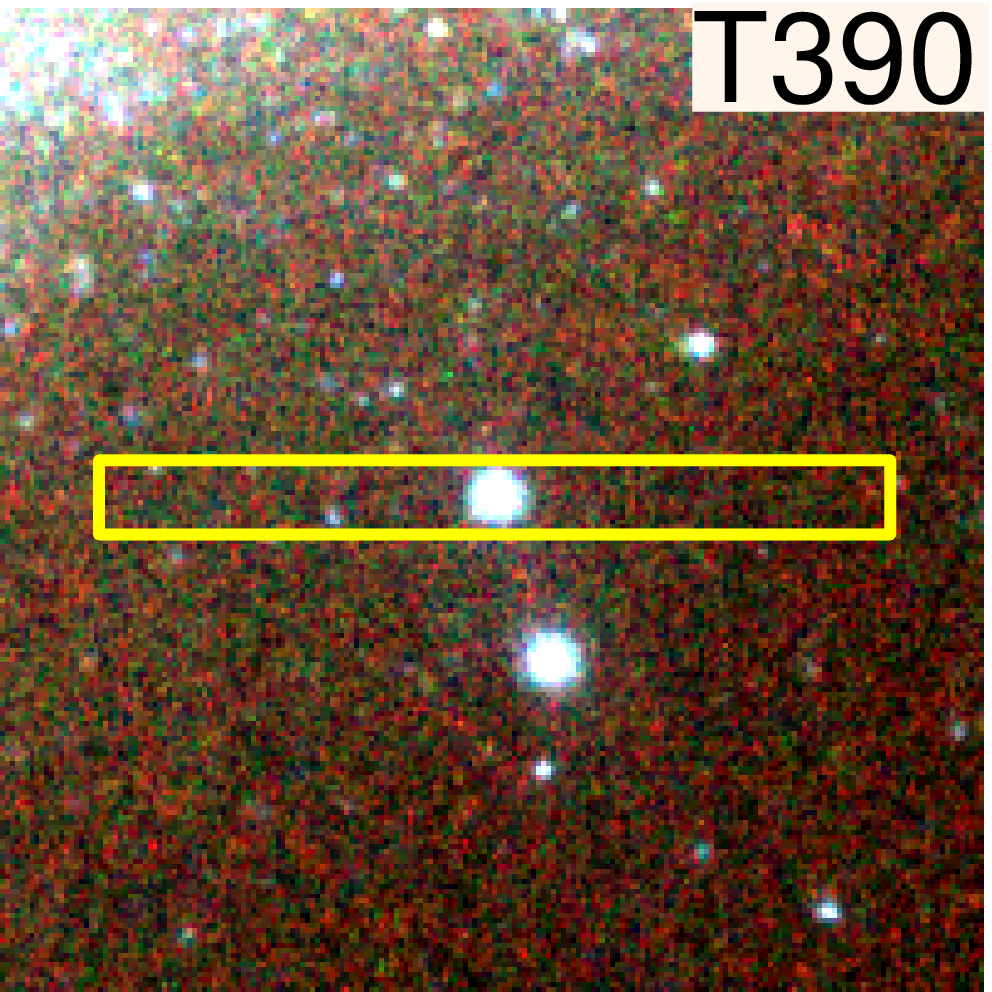}
\includegraphics[width=4cm]{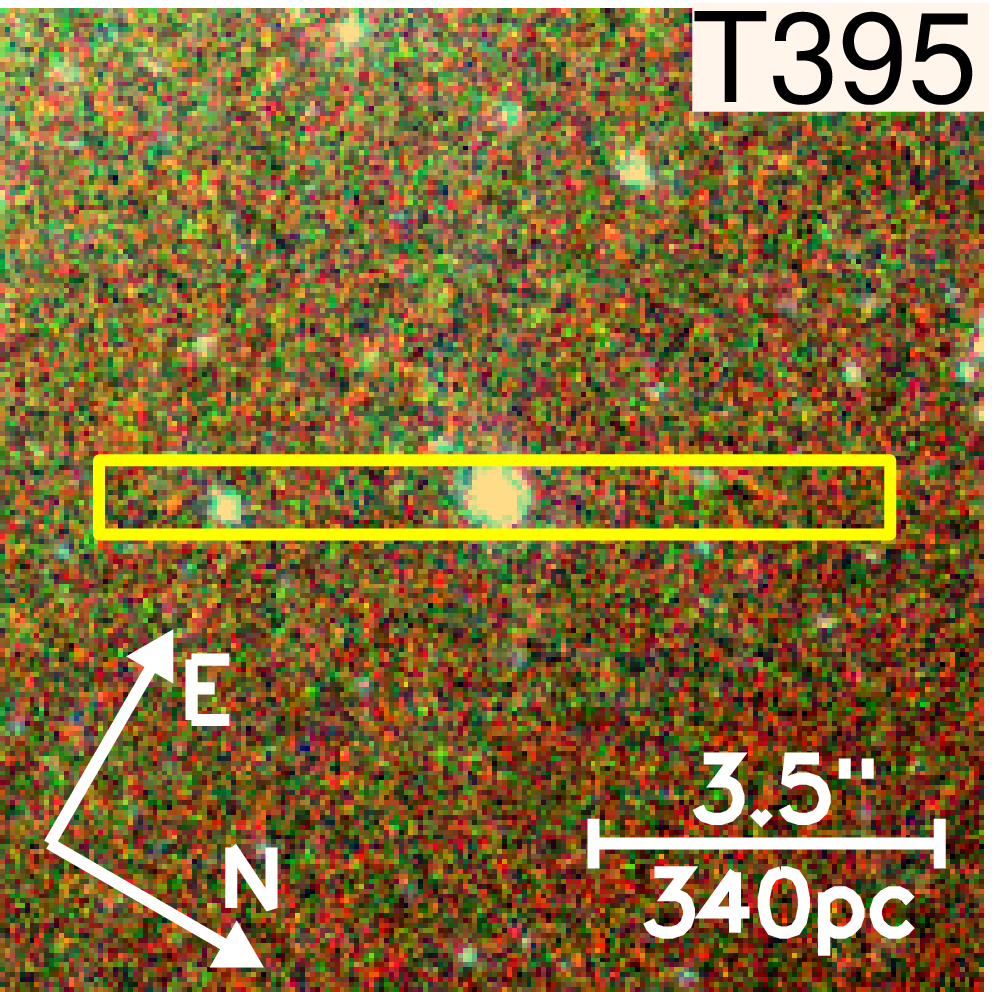}
\includegraphics[width=4cm]{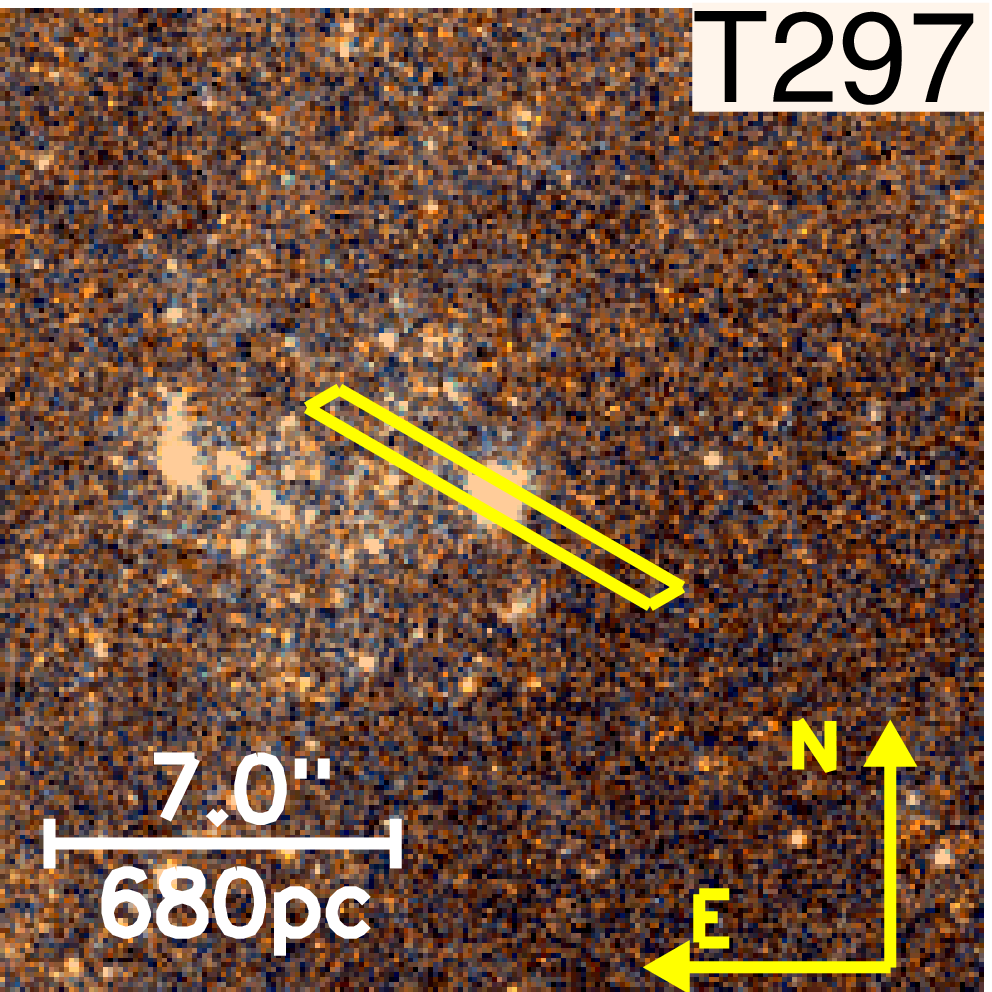}
      \caption{HST-ACS color images of each of the observed clusters or complexes. Each image is 200 pixels, or $\sim970$~pc on a side (except T297), and the GMOS slit is shown on each image.  Each image uses F435W, F550M, \& F658N for blue, green and red respectively, with the exception of T299 and T365 that use F435W, F814W and F658N (as the ACS chip gap appears in the F550M  field of view).  The image of T297 is made from WFPC2 F555W \& F184W exposures with a size of 200 pixels, or $\sim$1900~pc on a side.   All images, except that of T297, have the same scale and orientation.} 
         \label{fig:images}
      \end{center} 
%         \label{fig:images}
 \end{figure*}

In Fig.~\ref{fig:images} we show a color image centered on each of the observed clusters, with a field of view of 200~pixels ($\sim970$~pc) on a side, except for T297 where each side corresponds to $\sim1900$~pc.  The GMOS slit is shown in each image.  In Table~\ref{table:colors} we give the measured magnitudes for each cluster, where we have only corrected for Galactic extinction.

\begin{figure}
 \begin{center}
   \epsscale{1.15}
 \plotone{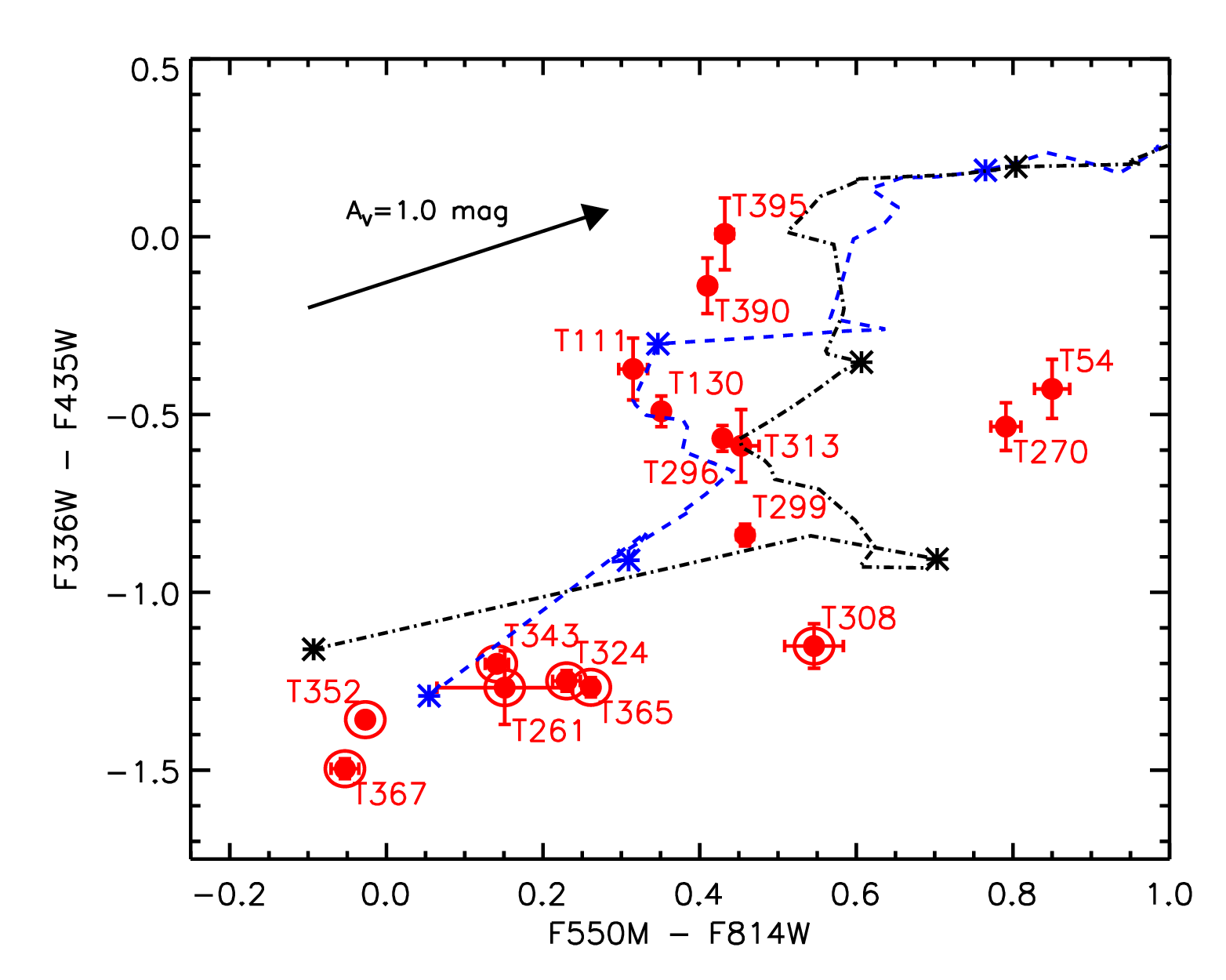}
      \caption{Color-color diagram of all clusters with available spectroscopy, measured from HST WFCP2 and ACS images.  Circled points indicate an excess of H$\alpha$ as derived from the HST imaging.  The dashed and dash-dotted lines represent  {\it GALEV} SSP models of half-solar and solar metallicity respectively.  Asterisks mark SSP ages of 4, 10, 100 and 1000~Myr (from bottom left to top right).  The reddening vector is shown for one magnitude of visual extinction. } 
         \label{fig:color-color}
      \end{center} 
 \end{figure}

The resulting color-color diagram is shown in Fig.~\ref{fig:color-color}.  We also show the {\it GALEV} simple stellar population models (Anders \& Fritze v. Alvensleben~2003) for solar (black - dot-dashed) and half-solar (blue - dashed) metallicity, that adopt a Salpeter~(1955) IMF and the Padova stellar isochrones.  The circled points represent clusters that display excess H$\alpha$ emission as derived through the HST photometry.  

\section{Derivation of Cluster Properties}
\label{sec:properties}
\subsection{Age and Metallicity}
\label{sec:ages}

\begin{figure*}
   \epsscale{1.00}
   	\plotone{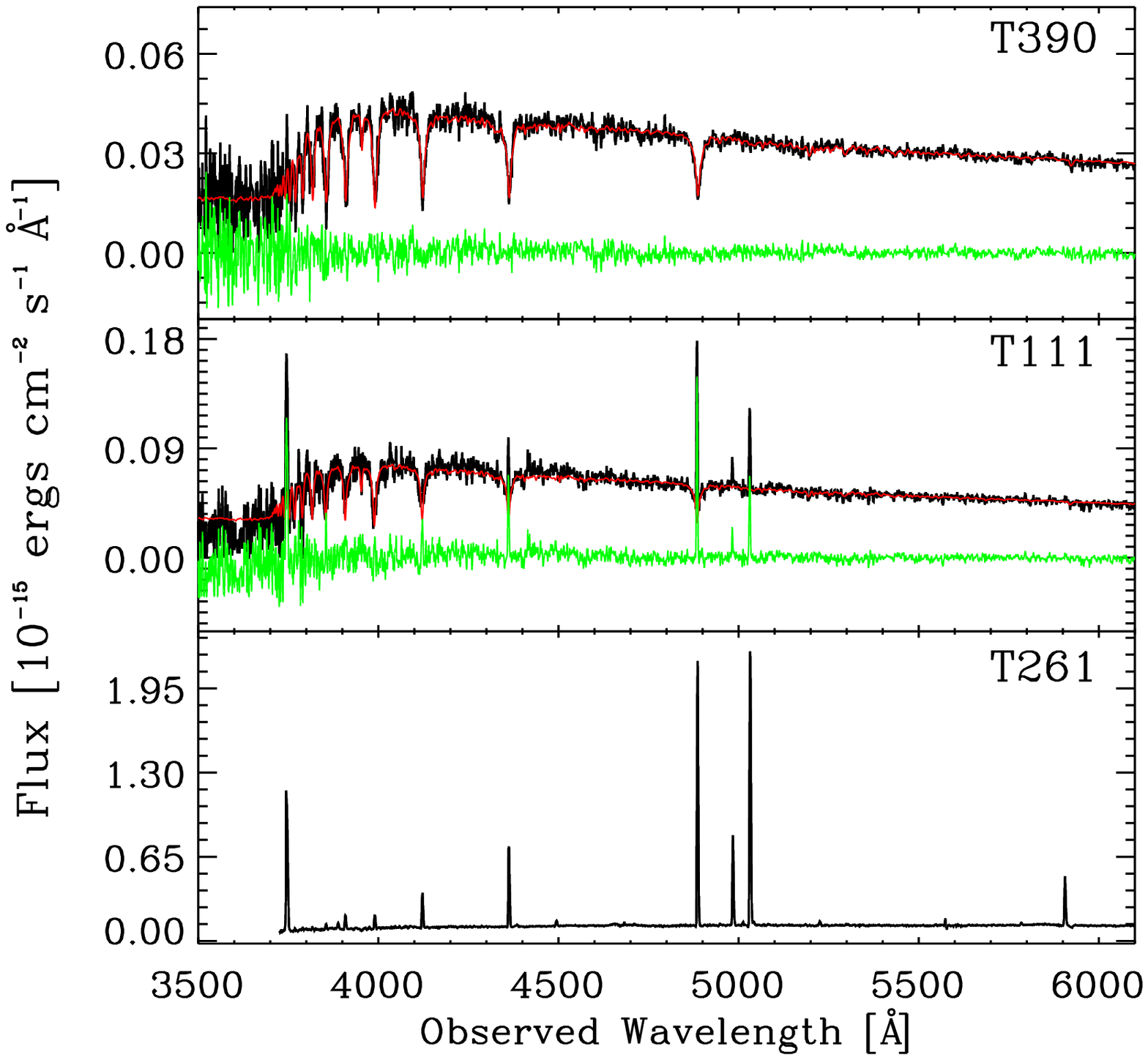}
       \caption{Example spectra of three clusters in our sample, corrected for the estimated
       interstellar extinction.  The red lines represent the best
       fitting (see \S~\ref{sec:ages} for a discussion of the method)
       model template.  The green lines represent
       the residual (best fitting template subtracted from the observed cluster).  The top panel shows a pure absorption line cluster, while the center panel shows a  $\sim80$~Myr old cluster that also shows emission features (presumably unrelated).  The bottom panel shows an example of a pure emission line cluster.}
   \label{fig:examples}
   \end{figure*}

The derivation of cluster properties (such as age and metallicity) based on the strengths of  stellar absorption lines through optical spectroscopy  is not a simple matter, due to degeneracies between age, metallicity, and extinction.  The observed spectra fall into two fairly distinct categories, absorption line dominated and emission line dominated.  In some cases, absorption and emission lines are seen superimposed.  The three cases are shown in detail in Fig.~\ref{fig:examples}.

%In some of our cluster spectra, the strengths of stellar absorption lines cannot be seen it or measured due to strong emission lines, but the equivalents width(EW) of the emission lines of the surrounding \hii region can be measured. 

\subsubsection{Absorption line clusters}

%We calculate the extinction of our spectra and we use the BC03 models, 

In order to estimate the age and metallicity of absorption line clusters, we measured the line strengths of a variety of Balmer and metal lines in each spectrum.  In T07b we outlined our technique in detail, and here we only give a brief summary of the method. 

We first construct a template for each cluster using the Penalized Pixel-Fitting (pPXF) method (Cappellari \& Emsellem 2004).  The method is based on the Bounded-Variables Least-Squares algorithm and constructs a template based on a combination of simple stellar population (SSP) models, and uses a penalized maximum likelihood formalism.  It has the advantage of being robust even when the data have a low signal-to-noise ratio (S/N).   The pPXF method, as realized in an IDL routine, takes in SSP model spectra of different ages/metallicities and weights them in order to create a best fit cluster template spectrum.  For the SSP models we chose the Gonz\'alez-Delgado et al.~(2005; hereafter GD05) models (of 1/5, 2/5, 1, and 2 times solar metallicity; Salpeter stellar IMF; and ages between 4~Myr and 15~Gyr) due to their high spectral and temporal resolution.

\begin{figure}
     \epsscale{1.15}
    	\plotone{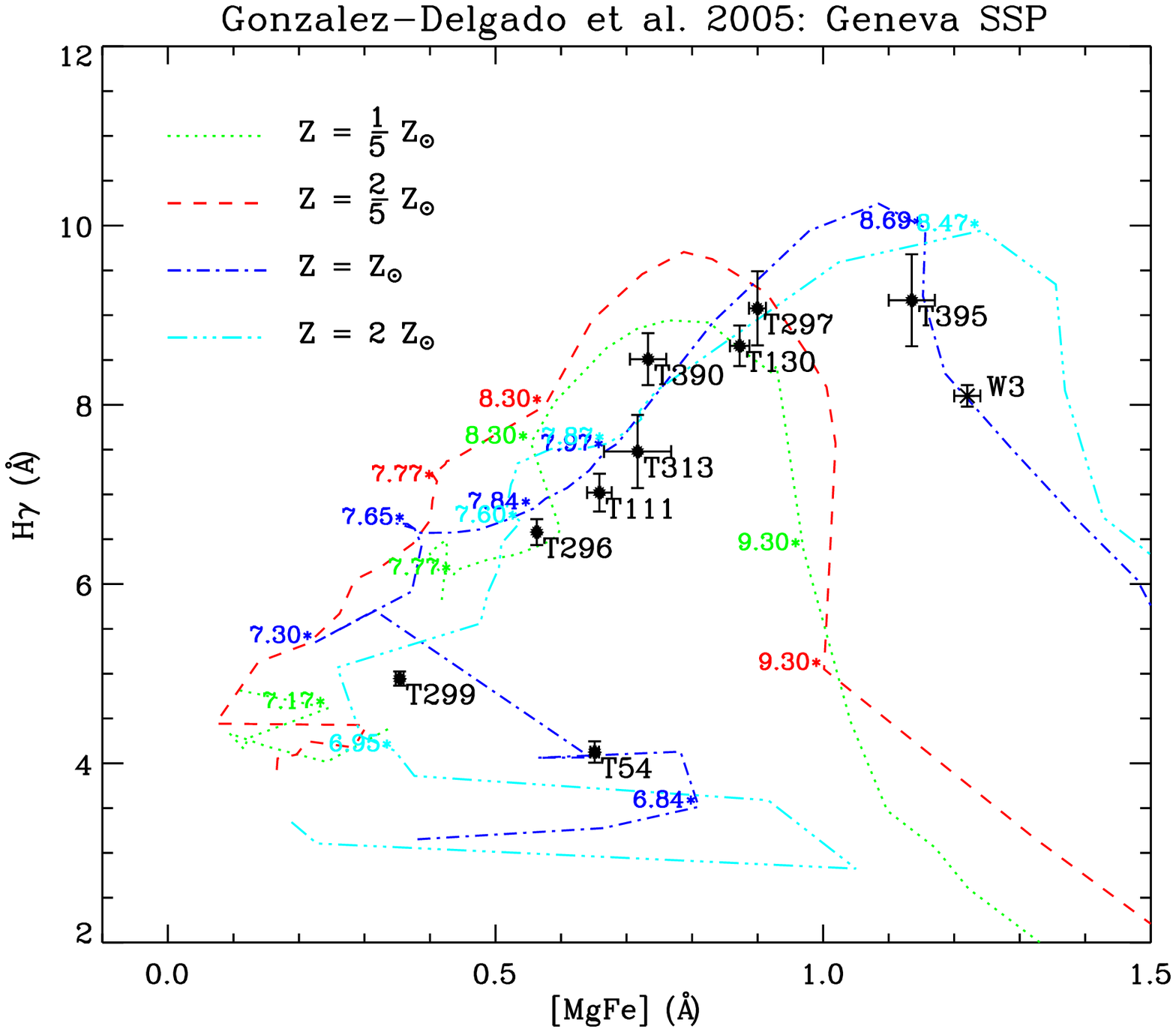}
      \caption{Determination of cluster ages and metallicities. H$\gamma$ vs. [MgFe] from the Gonz\'alez-Delgado et al.~(2005) SSP models for four different metallicities are shown.  Data points with error bars mark observed clusters and their  1-$\sigma$ errors.  In addition, we show the position of the
      massive cluster W3 in NGC~7252 for comparison. }
   \label{fig:index}
   \end{figure}

In order to estimate the ages of star clusters, we measured Lick line-strength indices (Faber et al.~1985;
Gonz\'alez~1993; Trager et al.~1998) and the indices defined by Schweizer \& Seitzer~(1998)  (i.e. HHe, K, H8)  from the output template spectra. The measurements of the indices were carried out with the task INDEX (Cardiel et al.~1998).   The errors on the measurements include the influence of photon-statistics and uncertainties in the radial velocities  and continuum level.  
In order to fully exploit the observations we compare all seven indices shown in Table~2 to the GD05 models weighted by their respective errors in a least $\chi^2$ sense. The values of the indices are given in Table~2.    In Fig.~\ref{fig:index} we show the resulting H$\gamma$ and [MgFe] indices for the clusters, along with SSP models.  For comparison we also show the position of W3, a massive cluster in the merger remnant NGC~7252 that has an age of $\sim500$~Myr and has approximately solar metallicity (Schweizer \& Seitzer~1998).

\begin{figure}
     \epsscale{1.15}
     \plotone{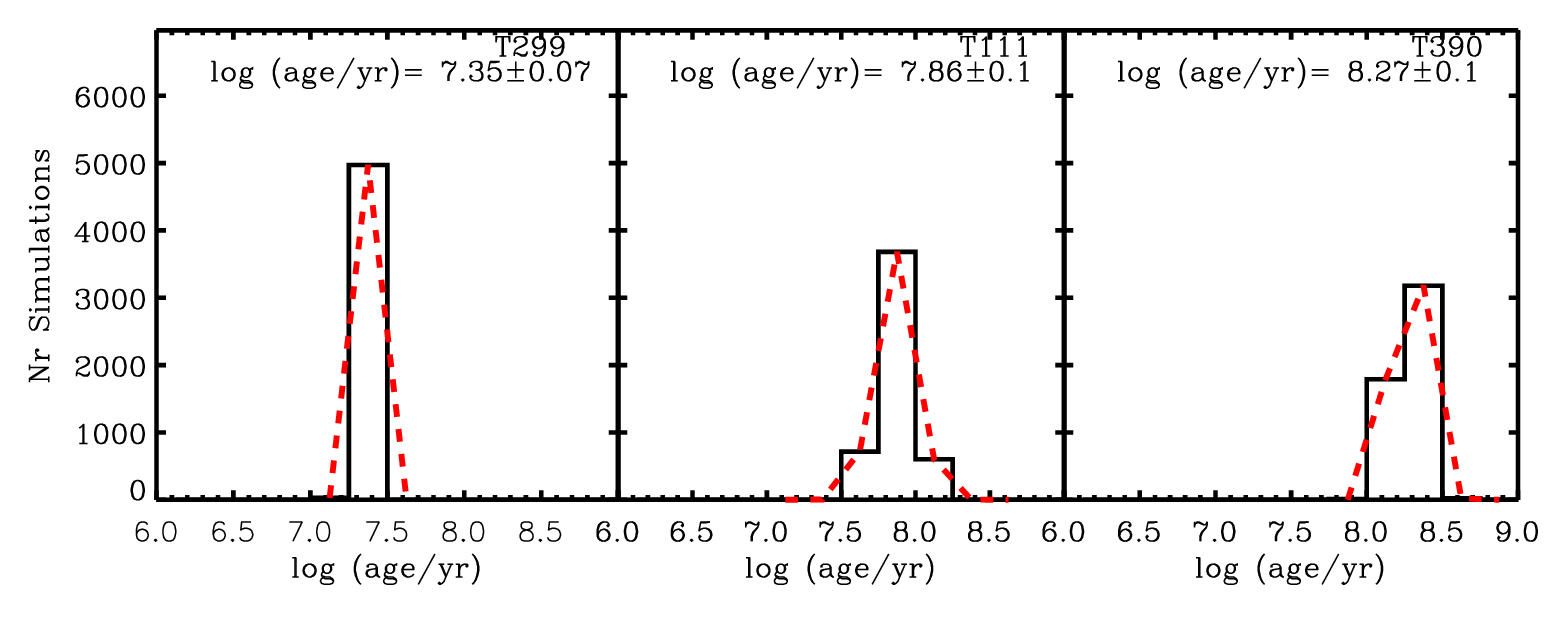}
     \plotone{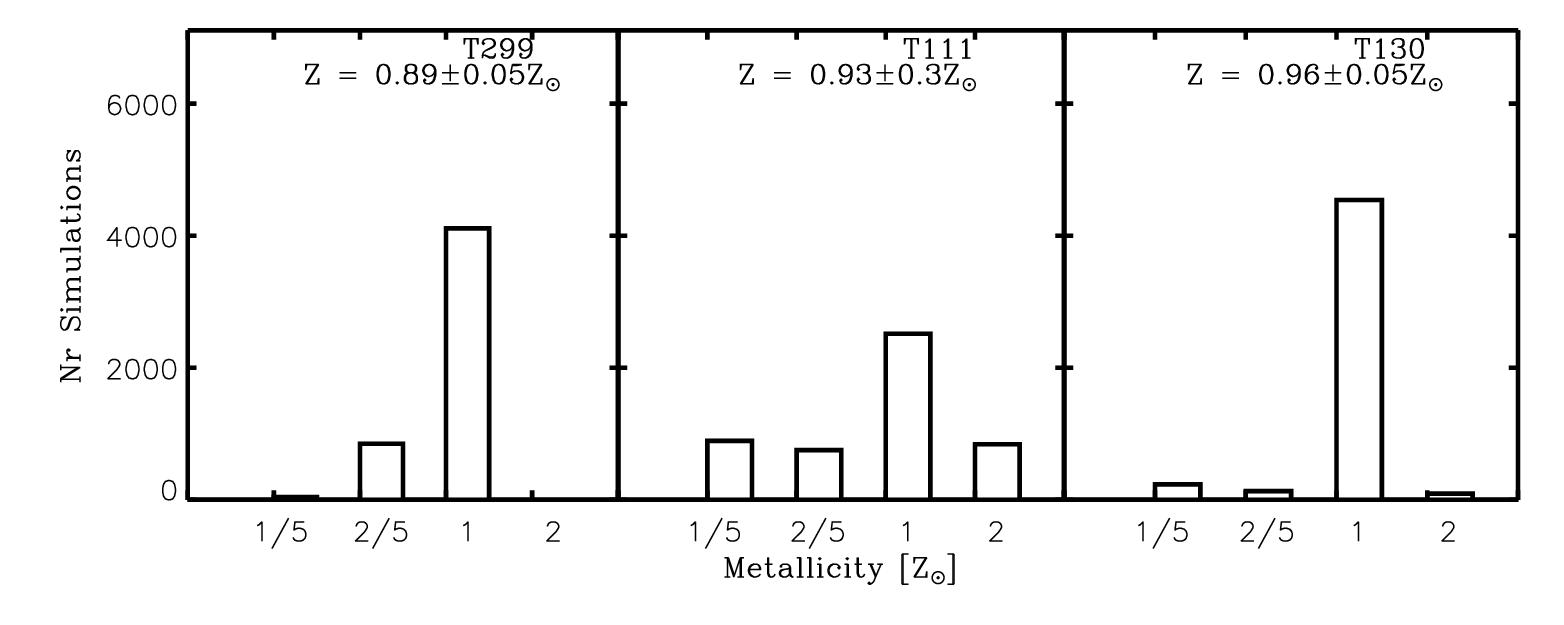}
    \caption{Probability distribution of the ages (top) and metallicities (bottom) derived by simulating the
     effect of the errors on the fitting routines.  See the text for details of the simulations.}
         \label{fig:age-hist}
 \end{figure}

The metallicities were derived in a similar way.     In order to estimate the errors on the age/metallicity measurements, we performed 5000 monte carlo simulations for each cluster, where we added noise to the measured indices sampled from the estimated errors.  We then fit the resulting age distribution with a Gaussian to estimate the final age and associated error and take the mean metallicity.  An example of the method is shown in Fig.~\ref{fig:age-hist}.

More details about this method are given in T07b.

\begin{figure}
     \epsscale{1.15}
     \plotone{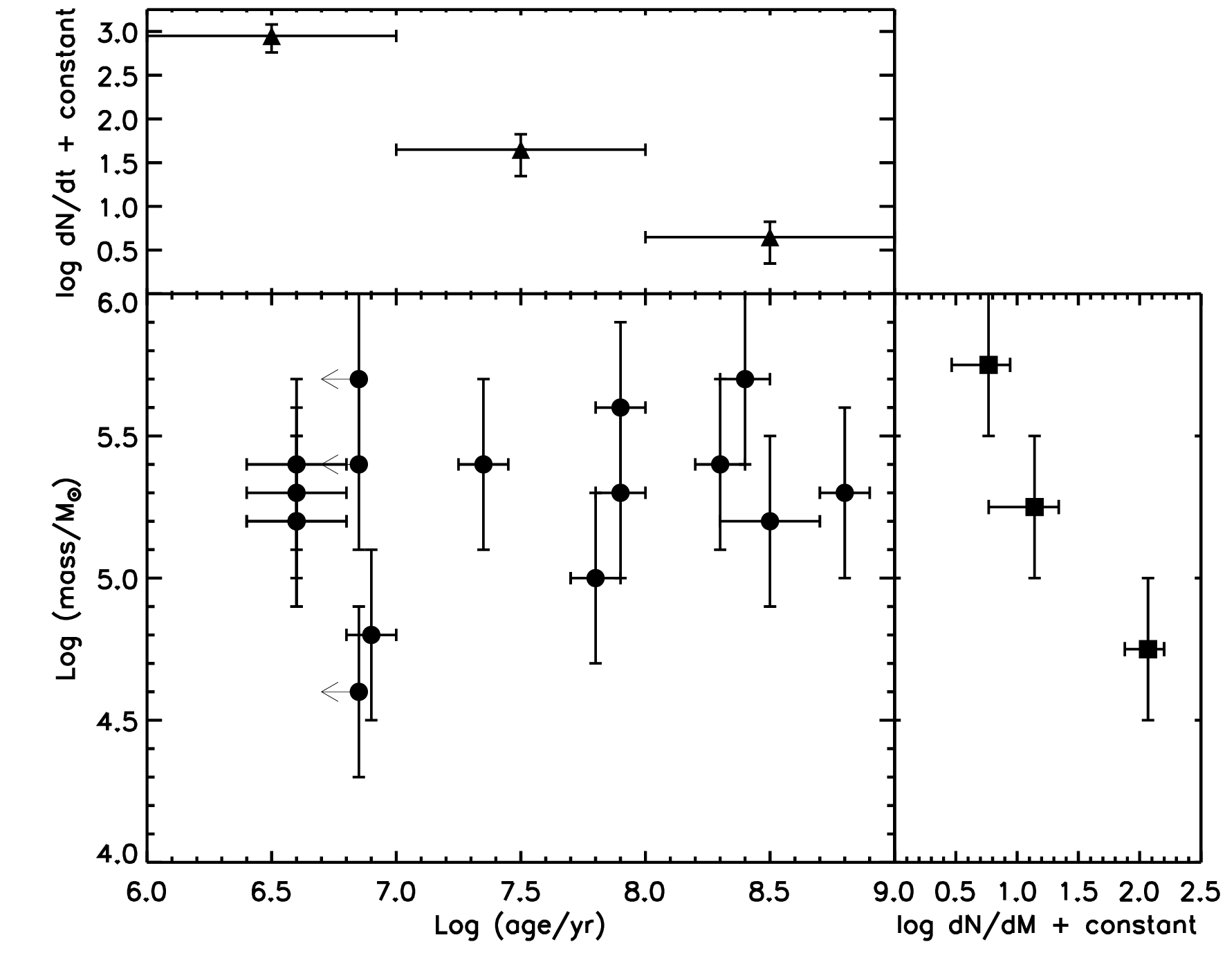}
    \caption{The age-mass diagram for the 16 clusters presented in this work (note that T324 \& T367 have the same estimated mass and age).  The top panel shows the age distribution (dN/dt) of the clusters while the right panel shows the mass distribution (dN/dm).  The horizontal "error bars" in the top panel and the vertical "error bars" in the right panel represent the binning used to create the distributions.  Poissonian errors are also shown in the age and mass distributions.  We caution that our sample is not complete in any sense. }
    \label{fig:age-mass}

 \end{figure}

	\subsubsection{Emission line clusters}

For the youngest clusters with little or no absorption features in their spectra the task of age dating is much easier.  First, we assign ages to these clusters of less than 10~Myr, due to the presence of large amounts of ionized gas around the cluster. Age dating can be refined to some degree by the presence or absence of Wolf-Rayet features, which, if present, restricts the age a range to $3-7$~Myr (see \S~\ref{sec:comparison}).

The metallicity of these clusters can be estimated from the surrounding nebular emission gas, which is likely to have similar (or the same) abundance as the young cluster.  Again, the method adopted here is the same as used in T07b, and we refer the reader to that work for details.

We adopt the chemical abundance analysis method from Kobulnicky \& Kewley (2004; hereafter KK04) to determine 
the metallicity. We measure the EW ratio of the collisionally excited  [OII]$\lambda$3727 and [OIII]$\lambda\lambda$4959,5007 emission lines relative to the H$\beta$ recombination line (known as R$_{23}$) and [OIII]$\lambda\lambda$4959,5007  relative to  [OII]$\lambda$3727  (knows as O$_{32}$), along with the calibrations of KK04 (their Fig. 7 - upper brach) and we use the solar abundance ratios by Edmunds \& Pagel (1984). Instead of the traditional flux ratio, the KK04 method uses EW ratios, which have the advantage of being reddening independent.  We note that strong line methods like those used here can be affected by the spatial distribution of the ionizing stars relative to the gas, as this can change the effective ionization parameter (Ercolano et al.~2007).  However, since detailed information on the spatial distribution and lack of detection of the much weaker auroral lines means that strong line methods are the only possibility in the present case.  With this caveat in mind, we list the derived metallicities in 
Table~\ref{table:colors}.

As can be seen,  the metallicites found for absorption-line and emission-line clusters agree well, giving us confidence in the robustness of the diagnostic methods and results.

\subsection{Velocities}

In the case of absorption-line clusters, we used the IRAF task
{\em rvsao.xcsao} for the determination of the redshift from the
individual spectra, using radial-velocity standard stars of three different types (HD~100953,
HD~126248, and HD~133955 of spectral types A, O, \& B respectively) observed at the same resolution as the clusters.
The three template stars were employed to reduce the
systematic errors introduced by the effect of template mismatch when
computing the redshift using the cross-correlation technique.

For the emission-line clusters, velocities were measured from the observed emission lines using the IRAF task {\em rvsao.emsao}.

In both cases the velocities were corrected to heliocentric velocity (see Table~\ref{table:properties2}).

%\section{Derivation of Cluster Properties}
%\label{sec:properties}

%\subsection{Age and Metallacity}
%\label{sec:ages}
	
%	\subsection{Absorption line clusters}
	
%	\subsection{Emission line clusters}
%Ercolano et al.~(2007) - possible problems with strong line emission measures.

%\subsection{Velocities}

\subsection{Extinctions and masses}

We noted a flattening in the continuum of the observed spectra that affected the data much like what would be expected by extinction.  However, based on the position of the clusters in color space (see Fig.~\ref{fig:color-color}), little or no extinction is expected in many (especially the older) clusters outside the main body of the galaxy (e.g.~T395).   The problem was traced to an excess of scattered light in the GMOS instrument on Gemini-North, which is particularly prominent with the B600 grating in the blue setting.   The result of this effect is that blue light is scattered to longer wavelengths causing the spectra of young clusters to appear flatter, thus mimicking the effect of extinction.  This should not have a significant effect on the measured equivalent widths nor the derived ages and metallicities.

Since the overall shape of the continuum is not precisely known, we determined the extinction based on the HST photometry. For each cluster we have derived its age using spectral features that are not affected by the overall continuum shape (\S~\ref{sec:ages}).  We then compared the measured UBVI colors of each cluster to SSP models of solar and half-solar metallicity of the appropriate age, and the observed colors were shifted along the extinction vector (assuming a Galactic extinction law of Savage \& Mathis~1979) until the closest point in color space is reached (see Konstantopoulos et al.~2009).  The derived extinctions were largely independent of the assumed metallicity, and are given in Table~\ref{table:colors}.

Once the age and extinction of each cluster was found the mass was calculated by comparing the extinction corrected luminosity to the mass-to-light ratio of the SSP models of the correct age.  For this we assumed a Kroupa~(2001) type stellar initial mass function that appears hold in massive extragalctic clusters (e.g. Maraston et al.~2004; Larsen et al.~2004; Bastian et al.~2006b; Goodwin \& Bastian~2006; Mengel et al.~2008). 

 The derived ages and masses are shown in the main panel of Fig.~\ref{fig:age-mass}.  We also show the resulting age (dN/dt) and mass (dN/dm) distributions, although we note that our sample is not complete in any sense.

\subsection{Sizes}

The size, expressed as the effective radius, of each of the clusters was measured from the HST-ACS F435W, F550M, and F814W images.  We used the {\it ISHAPE} package (Larsen~1999) that compares an analytic elliptical profile, convolved with a PSF, to the light distribution of each cluster.  We used a fitting radius of 10 pixels and PSFs made from stars in the globular cluster NGC~2419.  We also tested our size determinations against those of Mengel et al.~(2008) and find consistent results.  For three of the brightest clusters (with low surrounding backgrounds) we performed a fit using an Elson, Fall, and Freeman~(1987) profile.  This luminosity profile is expressed as a core with a power-law decrease in the outer regions.  In our fits for these clusters we allowed the index of the power-law to vary.  For these cases we carried out each fit with three initial guesses in order to validate that the best fit model was achieved, and not a local minimum in the $\chi^2$ distribution (see Larsen~2004).  In each of these cases, the best fitting power-law index was $\sim1.5$.   

We then used this profile to fit the remaining clusters, and as a test, also fit each cluster with an EFF profile with index 2.5, and King~(1962) profile with a concentration parameter of 30.  In general, we find consistent results with each of the profiles used, although the latter two tended to give slightly smaller effective radii.  The adopted effective radii are given in Table~\ref{table:properties2}.

While these clusters are slightly larger than average, compared to samples of young clusters in spirals (e.g. Larsen~2004; Scheepmaker et al.~2007), they are within the expected range.  This confirms that our sample is made up primarily of individual clusters.

\section{Results}
\label{sec:results}

\subsection{Comparison with previous work}
\label{sec:comparison}

Clusters T365 and T367 have been previously observed by Bastian et al.~(2006a; their complex 4 and 5 respectively) using the integral field unit (IFU) on VIMOS-VLT.  The authors noted strong Wolf-Rayet features in the spectrum, implying very young ages for these clusters ($3-7$~Myr), and used the observed [OIII] and H$\beta$ strengths to estimate the nebular metallicity, finding values slightly below solar.  In the present work, we have additional information available, namely the [OII]$\lambda3727$ emission line.  We find somewhat higher metallicities, being slightly super-solar.  We attribute this difference to the inclusion of the [OII] line and the refined measurement technique in the present work.  Our spectra also indicate the presence of Wolf-Rayet stars in the spectra of T365 and T367, confirming their young ages.

\subsection{Young clusters and cluster complexes}

As mentioned above, many of the young clusters, or cluster groups, show prominent Wolf-Rayet features in their spectra (see Fig.~\ref{fig:spectra});  the $\lambda4650$\AA\ "blue bump" and the $\lambda5800$\AA\ "orange bump" are seen in most of the emission line spectra.  The presence of this feature, formed in the outflows of massive stars reaching the end of their lives, indicates extremely young ages, between $3-7$~Myr (e.g. Sidoli et al.~2006 and references therein).  The ages indicated by the WR features agree with the photometric ages as inferred from Fig.~\ref{fig:color-color}, and also the study of Bastian et al.~(2006a).

We note that all but one of our emission line dominated spectra display WR features along with the nebular emission lines of H$\gamma$, H$\beta$, [OII] and [OIII].  From this we infer that nebular emission (at least on the size scale of our aperture, 0.75" or 72~pc) lasts for $\lesssim7$~Myr around massive clusters (i.e. if it lasted longer we would expect to see significant numbers of clusters with nebular emission but no WR features).  This is in agreement with the results of Whitmore \& Zhang~(2002) who found that the embedded phase of clusters in the Antennae must be quite short.

Additionally, we find that the youngest clusters (T261, T270, T324, T352, T365 \& T367) are not isolated, but rather are parts of extended star-forming regions, which also contribute to our spectra and photometry.  This is in agreement with the results of Zhang et al.~(2001) and Bastian et al.~(2006a) who found that clusters are often not isolated, but are part of an extended hierarchy of star-formation.  However, we note that, with the exception of T261 and T352, our extraction apertures are dominated by a single cluster. The exceptions are dominated by two clusters - (possibly the early stages of a binary cluster?).   The older clusters T54 ($\sim10$~Myr) and T299 ($\sim20$~Myr) appear to be largely devoid of extended associations suggesting that these regions are quickly lost to the background or fade beneath current detection limits, leaving only the most massive cluster visible.

\subsection{Velocities}
\label{sec:velocities}

In Table~\ref{table:properties2} we compare our measured velocities with the high resolution H{\sc i} map of Hibbard et al.~(2001).  The H{\sc i} velocity map clearly shows the rotation of the progenitor spirals.  The majority of the clusters in our spectroscopic sample follow the H{\sc i} rotation, hence we deduce that the clusters belong to the disks of the merging galaxies.    The exceptions being T297, T390 and T395, that lie just outside the main body and will be discussed in detail in \S~\ref{sec:297}. This is similar to what we found in the cluster population of NGC~3256  (although in this galaxy only a single galactic disk is observed): that the majority of the clusters are still associated with the rotating disk where they presumably formed.  

This is in contrast with older merger remnants, like NGC~7252 (Schweizer \& Seitzer~1998) and NGC~3921 (Schweizer, Seitzer \& Brodie~2004) whose clusters (those formed during the merger) exhibit "halo kinematics", i.e. orbits dominated by random motions rather than disk rotation.  This is to be expected as the Antennae merger has not progressed enough to have fully disrupted the progenitor spiral disks.  The implication is that the majority of cluster formation in mergers happens in a disk with the orbits becoming randomized as the merger progresses (i.e. once the nuclei coalesce).

\subsection{Clusters not associated with the galactic disks}
\label{sec:297}

\subsubsection{T390 and T395}

T390 and T395 are located northwest of the disk of NGC~4038.  These clusters appear to be different from the rest of the clusters observed in/near the galactic disks in that their velocities are significantly offset from the nearby H{\sc i} gas.   Thus, it appears that these clusters have "halo" kinematics,  by which we mean that the clusters do not belong to either of the progenitor disks.   Both clusters are quite massive ($2\times10^5$\msun) and relatively old ($200-500$~Myr).  If they no longer belong to the disk of the galaxy, as implied by their velocities, they are expected to be long lived.  Their ages imply that they formed during the first close passage of the progenitor spiral galaxies.%, which suggests that some cluster formation can occur outside the pre-existing disk.

\subsubsection{T297}

As part of our survey we placed slits on candidate objects in the tidal tails of the galaxy, as clusters have been recently found in these somewhat exotic environments (Gallagher et al.~2001; Knierman et al.~2003; Bastian et al.~2005; Knapp et al.~2006;  T07a,b; Chien et al.~2007; Werk et al.~2008).  One "bona fide" cluster was found, T297, the location of which is shown in Fig.~\ref{fig:image-tails} and a close up view of the cluster and surroundings is shown in Fig.~\ref{fig:images}.  Due to the position of this cluster in the MOS mask, the observed wavelength range was significantly redder than for the other clusters\footnote{The nature of MOS observations is such that if the slit is significantly offset from the center of the dispersion axis the resulting spectra will have a shifted wavelength range.}.  Therefore the template for this cluster was constructed primarily on the H$\beta$ and H$\alpha$ lines.  We then measured the resulting H$\gamma$ index from the constructed template in order to place T297 in Fig.~\ref{fig:index}. The age of the cluster puts its formation at the time of the first passage, roughly 200~Myr ago. 

 The projected position of the cluster implies that it is associated with  the southern tidal tail of the merging galaxies.  However, the velocity of this cluster is significantly offset from the nearby H{\sc i} gas.  This implies that this cluster is not physically associated with the southern tail that it is projected upon.  Deep CTIO 4m optical images of the Antennae (F. Schweizer priv. comm.) suggest that this cluster, and a few other nearby stellar associations, may be part of the extreme outer disk of NGC~4039.  The current position and velocity of the cluster implies that, regardless of its current state, it will eventually become part of the halo population of the remnant.  In this way, T297 is akin to T390 and T395 discussed above.

We note the presence of a stellar association near T297 showing colors very similar to that of the cluster (F555W-F814W$=0.6$).  Hence it is likely that these structures are physically associated.  We also note that these colors are consistent with the spectroscopic age determined for T297 ($\sim200$~Myr) with little or no extinction.  If this association is related to the cluster this would imply that they have remained spatially correlated for the past $\sim200$~Myr, which is similar to that seen in dwarf galaxies captured in the halo of the Milky Way (e.g.~Read et al.~2006).

\section{Discussion}
\label{sec:discussion}

With our relatively large sample of spectroscopically determined ages we can now investigate correlations with spatial position and implications for cluster formation/destruction scenarios and merger induced starbursts.  A deep understanding of these issues is essential if clusters are to be used to trace the major star forming episodes of galaxies.

\subsection{Age dependent positions}

One somewhat striking feature of the derived ages, is the presence of relatively old ($\sim200$~Myr) clusters outside  the main, bright star-forming part of the disks (e.g.~T130, T297, T313, T390, T395).  Part of this is likely to be a selection effect as some of these clusters would not be visible if they were located in the main body.  However, it does show that older clusters exist in this merger, with ages consistent with the time of the initial close passage.   Also, this shows that even at first periapse, the cluster formation was extended throughout the galaxy and not simply concentrated in the centers of the progenitor spirals.

The majority of the clusters (the exceptions being T297, T390, and T395)  have velocities consistent with the H{\sc i} velocity field (or its extension) as observed by Hibbard et al.~(2001).  The H{\sc i} gas has been previously shown to outline the rotation patterns of the progenitor disks, hence these clusters belong to the disks of NGC~4038/39.   We would expect the clusters to display "halo kinematics" as the merger progresses and the progenitor disks become destroyed, as seen in two intermediate age clusters in NGC~3921 (Schweizer et al.~2004).  As noted in \S~\ref{sec:velocities}, T297, T390 and T395 already display kinematics that are significantly offset from nearby H{\sc i} gas, despite their relatively young ages ($\sim200$~Myr).

\subsection{Age distribution and the cluster formation rate}

The age distribution of clusters has been widely used recently to estimate the star-formation history of galaxies (e.g.~Whitmore et al.~1999; de Grijs et al.~2003; Gieles et al.~2005; Smith et al.~2007; Konstantopoulos et al.~2009).  A complication of this method is that clusters do not survive indefinitely, but rather disrupt over some timescale, such that the age distribution of clusters is a combination of their formation and disruption history.  

FCW05 have studied the Antennae cluster population using HST imaging and have derived cluster ages and masses based on the comparison of the photometry to SSP models.  In this way they estimated the age of a few thousand clusters, a sample that is approximately an order of magnitude larger than most other studies to date.  Interestingly, they found a steep decrease in the number of clusters (for a mass limited sample) as a function of age.  In their analysis, they assumed that the cluster formation rate of the Antennae has remained constant for the past $1.5$~Gyr.  Based on this assumption, they interpreted the observed increasing (i.e. it is higher today than it was in the past) cluster formation rate as heavy cluster disruption, leading to 90\% of clusters dissolving every age dex and therefore a $\tau^{-1}$ age distribution.   This rapid decrease was interpreted as infant mortality\footnote{See Lamers~2008 for a comparison between two empirical disruption laws discussed in the literature.}.

Infant mortality is thought to be due to the removal of the residual gas of star-formation (arising from the $<100$\% star-formation efficiency).  While this phase is relatively short ($<$ few Myr), the process can leave a cluster far out of dynamical equilibrium for longer periods, leaving the cluster vulnerable to disruption.  Whitmore, Chandar, \& Fall (2007; hereafter WCF07) suggest that this process, along with stellar evolutionary mass loss, can result in a prolonged period ($\sim100$~Myr) of mass independent disruption, which we refer to as "long duration infant mortality".  Theoretical investigations, which do not include stellar evolution, suggest that this process should be largely over by $\lesssim10-20$~Myr (e.g. Goodwin \& Bastian~2006; Baumgardt \& Kroupa~2007 - see footnote 1).

Using the {\it STARBURST99} simple stellar population models (Leitherer et al.~1999) we can estimate the amount of mass loss expected due to stellar evolution.  Adopting a Kroupa~(2001) stellar IMF and solar metallicty for an instantaneous burst, a cluster is expected to lose 7\% of its mass in the first 10~Myr due to stellar evolution and a further 6-7\% in the subsequent 10 Myr (and a further 8-9\% from 20 to 100~Myr).  This relatively slow mass loss is not expected to significantly affect the evolution of a cluster unless the cluster is substantially mass segregated, which will tend to amplify the disturbance caused by stellar evolution (e.g. Vesperini et al.~2009).

%  The long duration mass independent disruption model was refined by Whitmore, Chandar, \& Fall~(2007; hereafter WCF07) who assume that the infant mortality phase lasts 100~Myr, still one order of magnitude longer than suggested from theory.}

 However, before searching for further causes of long duration mass independent disruption in order to explain the FCW05 and WCF07 observations, we note that other explanations exist.  In particular, it is highly uncertain whether the CFR has remained constant during this galactic merger.  Many recent works have provided evidence for an increased CFR during an interaction or merger (e.g. de Grijs et al.~2003; Konstantopoulos et al.~2008, 2009).  

Mihos et al.~(1993) have modeled the Antennae merger and found that the two progenitor spirals went through their closest passage $\sim200$~Myr ago.  Their simulations included a prescription for star-formation, and predict that the star-formation rate (SFR) should begin to increase at that time, with a current SFR more than six times higher than before the interaction began.  However, the models underestimate the SFR in the current interaction zone between the two galaxies, suggesting that a factor six increase is likely a lower limit.  { The current SFR of the Antennae is $\sim20$\msunyr\ (Zhang et al.~2001; although some estimates place this value as low as $\sim5$\msunyr - Knierman et al.~2003); assuming that the two progenitors were somewhat gas-poor spirals with SFRs of $1$\msunyr\ each, this results in an increase in the SFR of a factor of $\sim10$ since the merger began some $\sim200$~Myr ago.  Other, more recent simulations of mergers, which include various prescriptions for star-formation, predict the same general result: that the SFR abruptly increases at the time of closest passage and that the starburst lasts for a relatively short period of 200-500~Myr (e.g. Cox et al.~2008).

\subsubsection{Applying the predicted SFR increase to the observations}

Here we compare the observed age distribution of clusters with models of galactic mergers that include a prescription for star formation.   Once the age distribution is "corrected" for this increase we will compare the resulting distribution with cluster disruption models.%The difference between this and previous works is that we use external constraints, namely models for the increase of SFR in mergers and models for the duration of the infant mortality phase.

In the top panel of Fig.~\ref{fig:dndt} we show the age distribution (dN/dt - number of clusters formed/observed per Myr) from the observations of WCF07 (these observations are essentially the same as those presented in Fall~2004)\footnote{The data were taken from Fig.~4 of WC07.  We note that the y-axis label in their plot is incorrect and we have used a corrected version (B. Whitmore priv. com.).}.  Additionally, we show the time of first passage between the progenitor disks as a vertical dashed line.  The solid line shows the expected SFR increase of the models of Mihos et al.~(1993)\footnote{We assume that the cluster formation rate follows the SFR.  If the cluster formation efficiency increases with increasing SFR (e.g.~Zepf et al.~1999) then the increase would be even higher than that shown.}.  The y-scaling was chosen to best match the two oldest observed age bins.  Finally, we show the age range over which infant mortality is expected to act.%Under the assumption that the cluster formation rate (CFR) began to increase at the time of closest passage and continued until $\sim10$~Myr, we show the expected dN/dt for total increase of five (squares) or ten (circles) times the initial CFR.  The begining and end of the increase are shown at log age = 8.3 and 7.0, respectively.  We connect the points using a straight line for an illustrative example, which implies a logarithmic increase.  

As can be seen, the observed data for ages greater than 10~Myr can be explained by the models very well.  This is explicitly shown in the bottom panel of Fig.~\ref{fig:dndt} where we have subtracted the SFR models of Mihos et al.~(1993) from the observations.  The horizontal dashed line represents an ideal fit between model and observations.  The fact that the observed data points are now all consistent with a flat distribution for ages larger than 10~Myr implies that no long duration (i.e. $>10$~Myr) infant mortality is required to explain the data.  The only data that deviate from this accord (i.e. more clusters per Myr are observed than predicted by the models) have ages less than 10~Myr.  This is consistent with the expected duration of infant mortality calculated using N-body and analytic models (e.g. Goodwin \& Bastian~2006; Baumgardt \& Kroupa~2007) and is also in agreement with observations of the star cluster population of the SMC (Gieles et al.~2007; de Grijs \& Goodwin~2008).

 We note that in the case of mass dependent cluster disruption, the dN/dt of a cluster population is largely expected to follow the SFR.  This is because the low mass (faint) clusters, that are preferentially destroyed, are generally below the detection limit (or mass cut) and hence do not contribute to the observed cluster age distribution.

We conclude that models of galactic mergers that include star formation can reproduce the observed age distribution of clusters. The only caveat to this arises in the youngest ages ($<10$~Myr), as infant mortality can affect this period.  Hence, the observed age distribution of clusters in the Antennae galaxies can be explained within the bounds of existing models, without the need of invoking long duration mass independent cluster disruption, which would require its own new theoretical framework.

\subsubsection{Other considerations of cluster disruption}

There is some additional evidence for a low (mass independent) cluster disruption rate in clusters that have survived the transition from the embedded to the exposed phase.  FCW05 show the age distribution of a {\it luminosity limited} sample.  This shows a similar behavior as the mass limited sample, decreasing as $\tau^{-1.25}$.  We note that this is inconsistent with the interpretation of long duration infant mortality.  In such a case, the cluster population will decrease due to a combination of disruption and evolutionary fading.  This results in a declining age distribution where the index (assuming that each follows a power-law form) is the sum of the disruption law and the fading curve (Gieles~2008, Konstantopoulos et al. in prep), the latter can be estimated from SSP models.  For the V-band, the index of the fading curve is $\sim0.7$ (Gieles~2008)\footnote{This is the same as $\zeta$ introduced by Boutloukos \& Lamers~(2003).  In fact the expected decrease is $\zeta(1-\alpha$) where $-\alpha$ is the index of the mass function (for a pure power-law MF).  If the underlying mass distribution is described by a Schechter function, then $\alpha$ is the index of the mass function at the luminosity being studied, which, in this case, would mean  $\alpha > 2$.  This means that the expected decrease, just due to fading, would be $<-0.7$, i.e. exacerbating the disagreement between the model of FCW05 and the observations.  See Gieles~(2009) for a full description.}
%\fbox{look at Boutloukos \& Lamers 2003, and Marks SMC paper - maybe a footnote to saying that it is also related to zeta*alpha - so all fucked up with a schechter function}, 
hence to be consistent with 90\% cluster disruption per age dex, the observed luminosity limited sample would need to decrease as $\tau^{-1.7}$, i.e. much steeper than observed.

Other evidence against long duration infant mortality was provided by Mengel et al.~(2008), who measured the dynamical masses of 9 clusters in the Antennae with ages between 6 and 9~Myr.  By comparing their dynamical and photometric mass estimates to models of early cluster evolution (Goodwin \& Bastian~2006) they found that all but one of these clusters are stable and bound.  This is in agreement with the models of Goodwin \& Bastian~(2006) and Baumgardt \& Kroupa~(2007) who found that if a cluster survives the transition from embedded to exposed states it quickly reaches equilibrium.  If long term disruption due to the removal of gas was taking place in the Antennae (FCW05) then it would be expected statistically that all but one would be out of equilibrium, i.e. the opposite of what is actually observed.

We also note that such long duration (mass independent) disruption is not favored in the comparison of observations of the brightest (and 5th brightest) clusters in a sample of 27 spiral and dwarf galaxies with the expectations of cluster population models (Larsen~2009).  This sample, in particular the spiral galaxy component, avoids the assumptions/problems mentioned above, as the SFR in these galaxies is expected to be relatively constant over the period of a few hundred Myr.

The masses of the observed clusters (see Table~\ref{table:properties2}), when compared with the results of N-body (Baumgardt \& Makino~2003) and analytic (Lamers et al.~2005) predictions of cluster disruption, suggest that many are expected to live to old ages ($>1$ Gyr).  Hence these observations agree with models that predict that gas rich mergers can produce significant numbers of metal rich clusters, which will be seen around the merger remnant, long after star-formation ceases (e.g. Schweizer~1987; Ashman \& Zepf~1992).

\begin{figure}
     \epsscale{1.0}
    	\plotone{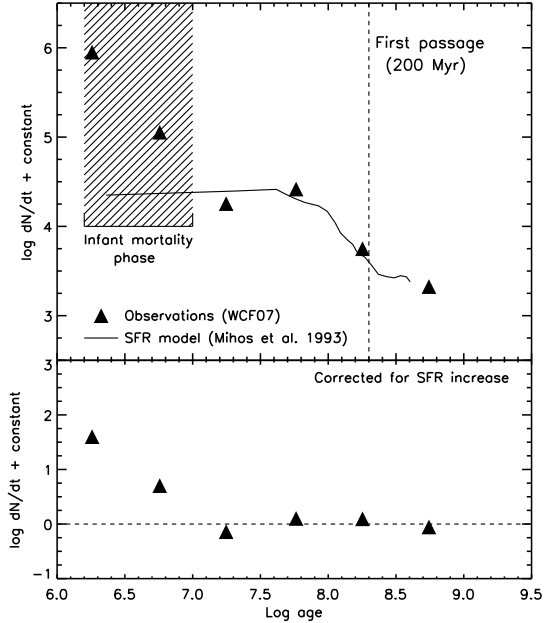}
      \caption{{\bf Top:} The observed age distribution (triangles) of clusters in the Antennae from WCF07.   The dashed vertical line shows the time of the first passage of the progenitor spirals.  The solid line shows the expected increase in the SFR due to the merging of the progenitor spirals from the models of Mihos et al.~(1993).  The shaded region is the time that "infant mortality" is expected to operate based on theory.   {\bf Bottom:} The observed age distribution when corrected for the SFR increase of Mihos et al.~(1993).  The dashed horizontal line indicates a perfect agreement between observations and the models.  Note that there is still evidence of "infant mortality" in the first $\sim10$~Myr and then the distribution becomes largely flat.  This shows that long duration mass independent cluster disruption is not required to explain the observations. }
   \label{fig:dndt}
   \end{figure}

\subsection{Comparison of the Antennae with other major mergers}

The masses derived for the clusters in the Antennae are generally lower than the ones derived in our survey of clusters in NGC~3256 (T07a,b).  This is partly due to a selection effect, as the distance to the Antennae is approximately half that than to NGC~3256.  In the later case we could only observe the brightest clusters, meaning we were biased to the higher mass clusters.  However, this may also reflect a physical difference between the systems, as the SFR of the Antennae ($\sim20$\msunyr) is less than half that  of the NGC~3256 system ($\sim50$\msunyr - Bastian~2008).  This difference in the SFR of the two galaxies is likely a combined effect of orbital/encounter geometry and gas distribution in the progenitor spirals (e.g. Cox et al.~2008).  NGC~3256 would then be expected to form clusters more massive than the ones in the Antennae, simply due to size-of-sample effects (e.g~Bastian~2008).  

In Fig.~\ref{fig:zhist} we show the derived metallicity distribution of clusters in the Antennae and NGC~3256 (both emission and absorption line dominated clusters).  Both systems are producing metal rich clusters, although NGC~3256 appears to be slightly more metal rich.  This is in agreement with findings for intermediate age mergers, where the cluster population that formed during the merger has roughly solar metallicity (Schweizer \& Seitzer~1998; Goudfrooij et al.~2001).

\begin{figure}
 \begin{center}
   \epsscale{1.}
 \plotone{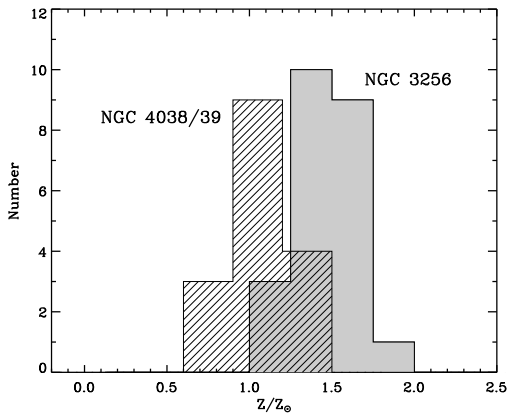}
      \caption{The distribution of the derived cluster metallicities (including both emission and absorption line clusters) in the Antennae (this work) and NGC~3256 (T07b).} 
         \label{fig:zhist}
      \end{center} 
 \end{figure}

\subsection{Merger induced star-formation models} 
\label{sec:induced}

The ages that we derive for clusters in the Antennae are broadly consistent with those found in other merging/interacting galaxies (NGC~3256 - T07a,b; NGC~4676 - Chien et al.~2007);  in that we find some clusters whose ages agree with the first passage of the galaxies and many young clusters associated with the recent burst.  The extended distribution of the young clusters within the galaxy is in accord with the shock-induced star-formation models, as opposed to pure gas density thresholds (Barnes ~2004).  This is somewhat surprising given the finding of Whitmore et al.~(2005) of small velocity differences ($<10$~km/s) among the cites where the formation of young clusters is taking place.  This rules out high velocity cloud-cloud collisions as the trigger of cluster formation.

%between young cluster formation sites is quite low ($<10$~km/s) ruling out high velocity cloud-cloud collisions as the trigger of cluster formation.

\section{Conclusions}
\label{sec:conclusions}

We have presented optical spectroscopy (complemented with HST imaging) of 16 star clusters in the merging Antennae galaxies.  This forms part of a larger spectroscopic survey of star clusters in merging/interacting galaxies across the Toomre sequence.  We have derived their ages, metallicities, masses and extinctions by comparing the observed spectra and photometry to simple stellar population models.  Our main results are as follows:

\begin{itemize}

%\item We find clusters with ages between a few Myr and 200~Myr.  This is consistent with the galactic merger models that include star-formation of Mihos et al.~(1993), who predict that the SFR of the Antennae has increased from the time of the first peri-center passage ($\sim200$~Myr ago) to the present time.

\item We find clusters with ages between a few~ÊMyr and $\sim600$~Myr, although all but three of the clusters have ages $<200$~Myr. This is consistent with the results of Mihos et al.~(1993), who modeled the interaction history of the Antennae galaxies and included a prescription for star-formation.  These models~ÊpredictÊ that the SFR of the Antennae has increased from the time of the first peri-center~ÊpassageÊ($\sim200$ Myr ago) to the current era by a factor of $>6$.

\item The metallicities estimated for the absorption and emission line clusters agree well with an average value of roughly solar.  This is consistent with models that predict gas rich mergers can produce a significant number of metal rich star clusters.  Based on the estimated masses of the clusters, many of them are expected to survive past a Gyr.

\item Most of the clusters observed in our sample are kinematically associated with one of the two disk galaxies involved in the merger.  This is similar to what was found by T07b for NGC~3256, a merger in a more advanced stage, where 14 out of 23 clusters in the main body follow disk rotation.  This implies that the majority of cluster formation in mergers happens in a galactic disk with the orbits becoming randomized as the merger progresses.

\item Three clusters, T297, T390 and T395, which lie off the disk of NGC~4038 in projection, have significantly different radial velocities than the nearby gas.  This implies that they do not belong to the disk components of the galaxies, but rather belong to the future halo.  Their ages place their formation concurrent with the first close passage of the progenitor galaxies.  Their ages, masses, and locations imply that these clusters are likely to be long lived.

\item  T297 is particularly interesting, as it appears projected onto the southern tidal tail of the Antennae.   However, its velocity is significantly offset from the H{\sc i} gas, implying that it is not physically associated with the tail, but rather belongs to either the halo or the extreme outer disk of NGC~4039.  We also discovered a stellar association around the cluster, with similar $V-I$ colors, suggesting the same age as the cluster.  If it is physically associated this would imply that material formed in this event (i.e. the cluster and the association) has remained spatially coherent for the past $\sim200$~Myr.

\item We have shown that the assumption of a constant star/cluster formation rate for the Antennae merging galaxies for the past 1~Gyr can lead to an overestimation of the fraction of dissolving clusters.   Comparing the present observational properties of the Antennae and its cluster population to models of the expected increase of the SFR due to the galactic merger (Mihos et al.~1993), we find a good general agreement.  Hence, we conclude that models of galactic mergers that include star formation can reproduce the overall age distribution of clusters well.   The only caveat to this general agreement is in the first 10~Myr of a cluster's existence, as infant mortality can affect this period.  We note that the galactic merger models used here make a number of simplifying assumptions (e.g. lack of shock induced star-formation) and therefore  it would be insightful to model this merger with state-of-the-art codes and techniques along with improved orbital modeling.

\item Once we account for the increase of the SFR due to the merger, we do not find any evidence for mass independent long duration cluster disruption (i.e. infant mortality lasting $>10$~Myr).  This is in better agreement with N-body and analytic considerations of the "infant mortality" phase.  Hence, the observed age distribution can be explained within the bounds of existing models without the need of invoking long duration mass independent cluster disruption, which would require its own new theoretical framework.

\item Based on the above, we suggest that while galaxy mergers such as the Antennae offer large samples of clusters, uncertainties in the cluster formation history and any possible changes in the cluster formation efficiency, limit their use to constrain cluster disruption.  Instead, we suggest that studies on cluster formation and evolution are best done in more quiescent environments where the assumption of a constant cluster formation rate and efficiency is more readily justified.  The results of these studies can then be applied to cluster populations within mergers in order to understand their star formation histories.

\end{itemize}

\begin{acknowledgements}
We thank Mark Gieles and TJ Cox for insightful discussions on cluster disruption and the star-formation history of mergers, respectively.  We are grateful to the referee, Fran\c{c}ois Schweizer, for helpful comments that improved the manuscript and interpretation of the data.  We also thank Linda Smith for comments and suggestions on the munuscript.  GT thanks Matt Mountain and Phil Puxley for the tremendous support during all these years.  NB is supported by an STFC fellowship and gratefully acknowledges Gemini Observatory for generous travel support.  ISK was supported by a Gemini Observatory grant.

Based on observations obtained at the Gemini Observatory (GN-2003A-Q-33), which is operated by the Association of Universities for Research in Astronomy, Inc., under a cooperative agreement with the NSF on behalf of the Gemini partnership: the National Science Foundation (United States), the Science and Technology Facilities Council (United Kingdom), the National Research Council (Canada), CONICYT (Chile), the Australian Research Council (Australia), MinistŽrio da Cincia e Tecnologia (Brazil) and Ministerio de Ciencia, Tecnolog'a e Innovaci—n Productiva  (Argentina)

%Based on observations obtained at the Gemini Observatory, which is operated by the Association of Universities for Research in Astronomy, Inc., under a cooperative agreement with the NSF on behalf of the Gemini partnership: the National Science Foundation (United States), the Particle Physics and Astronomy Research Council (United Kingdom), the National Research Council (Canada), CONICYT (Chile), the Australian Research Council (Australia), CNPq (Brazil) and CONICET (Argentina)
\end{acknowledgements}

%######################Table4###############################

%%%%%%%%%%%%%%%%%%%%%%%  Table 3' %%%%%%%%%%%%%%%%%%%%%%%%

%%%%table3%%%%%

\begin{deluxetable}{lccccccc}
\def\psn{\phs\phn}
\def\pnn{\phn\phn}
\tablecolumns{8}
%\tablenum{3}
\tablewidth{0pt}
\tablecaption{Spectral Line Indices}
\tablehead{
\colhead{ID} &\colhead{ $H+H\epsilon$\tablenotemark{a} } &\colhead{ $K$\tablenotemark{a} } &\colhead{ $H8 $\tablenotemark{a} }  &\colhead{$H\gamma_A$\tablenotemark{b}}    &\colhead{ $Mgb5177$\tablenotemark{b}}    &\colhead{ $Fe5270$\tablenotemark{b}}  &\colhead{ $Fe5335$\tablenotemark{b}} \\
\colhead{} & \colhead{ (\AA) } & \colhead{ (\AA) } & \colhead{ (\AA) }  & \colhead{( \AA)}& \colhead{( \AA)}& \colhead{( \AA)}& \colhead{( \AA)} 
}
\startdata
T54      & 5.18$\pm$0.19 & 0.75$\pm$0.09 & 3.26$\pm$0.19 &  4.12$\pm$0.11 &  0.42$\pm$0.07 & 0.90$\pm$0.08 & 1.21$\pm$0.12\\
T111    & 7.60$\pm$0.29 & 0.91$\pm$0.17 & 6.71$\pm$0.30 &  7.02$\pm$0.21 &  0.37$\pm$0.11 & 1.02$\pm$0.14 & 1.48$\pm$0.22\\ 
T130    & 9.83$\pm$0.31 & 0.76$\pm$0.18 & 8.73$\pm$0.31 &  8.65$\pm$0.22 &  0.64$\pm$0.12 & 1.05$\pm$0.15 & 1.46$\pm$0.22\\
T296    & 7.02$\pm$0.19 & 0.77$\pm$0.01 & 6.10$\pm$0.20 &  6.57$\pm$0.14 &  0.30$\pm$0.08 & 0.96$\pm$0.00 & 1.23$\pm$0.06\\
T297    &...                          & ...                          & ...                         &  9.07$\pm$0.41 &  0.73$\pm$0.15 & 1.00$\pm$0.07 & 1.36$\pm$0.23\\ 
%T299    & 7.14$\pm$0.10 & 0.77$\pm$0.06 & 6.16$\pm$0.11 &  6.66$\pm$0.08 &  0.30$\pm$0.04 & 0.97$\pm$0.06 & 1.24$\pm$0.09\\
T299    & 5.88$\pm$0.11 & 0.77$\pm$0.06 & 4.70$\pm$0.11 &  4.94$\pm$0.08 &0.20$\pm$0.04 & 0.57$\pm$0.06 & 0.67$\pm$0.09\\
T313    & 7.48$\pm$0.25 & 0.71$\pm$0.04 & 7.00$\pm$0.61 &  7.47$\pm$0.40 &  0.44$\pm$0.22 & 1.02$\pm$0.27 & 1.51$\pm$0.21\\ 
T390    & 9.43$\pm$0.43 & 0.72$\pm$0.25 & 8.35$\pm$0.45 &  8.50$\pm$0.29 &  0.45$\pm$0.15 & 1.08$\pm$0.19 & 1.46$\pm$0.28\\
T395   &11.20$\pm$0.72 & 2.97$\pm$0.41 & 9.94$\pm$0.78 &  9.16$\pm$0.51 &  0.77$\pm$0.21 & 1.58$\pm$0.26 & 1.86$\pm$0.37\\ 

\enddata
\tablenotetext{a}{\,Index definition by Schweizer \& Seitzer (1998).}
\tablenotetext{b}{\,Lick indices.}
\label{table:indices}
\end{deluxetable}

%%%%%%%%%  End of Tables  %%%%%%%%%

%######################Table4###############################

\begin{deluxetable}{cccccccc}
\tablecolumns{8}
\tablecaption{Kinematics and Masses of the clusters.}
%\tablenum{4}
\tablewidth{0pt}

\tablehead{
\colhead{ID} & \colhead{Agreement\tablenotemark{a}}  & \colhead{cz(HI)\tablenotemark{b}} & \colhead{czhel} & \colhead{deltcz}  &  \colhead{log(Mass)}  & \colhead{R$_{\rm eff}$}\\ 
\colhead{} & \colhead{} & \colhead{(km/s)} & \colhead{(km/s)} & \colhead{(km/s)} & \colhead{\msun} & \colhead{(pc)}} 

%% All data must appear between the \startdata and \enddata commands
\startdata
T54     &0  & 1700    & 1697$\pm$54  &-3     &4.8$\pm$0.3  & 3.7 \\  
T111    &0  & 1560    & 1595$\pm$115  &+35    &5.3$\pm$0.3   &  6.7 \\  
T130    &0  & 1565    & 1617$\pm$61  &+52    &5.7$\pm$0.3   &   6.0 \\  
T261    &0  & 1670    & 1621$\pm$13  &-49    &4.6$\pm$0.3  &     -- \\
T270    &0  & 1715    & 1711$\pm$19  &-4     &5.4$\pm$0.3   &    9.3\\
T296    &0  & 1755    & 1733$\pm$35  &-22    &5.6$\pm$0.3   &     4.0\\  
T297    &1  & 1675    & 1553$\pm$41  &-122    &5.2$\pm$0.3   &     --\\
T299    &0  & 1795:\tablenotemark{c}   & 1810$\pm$38  &+15:   &5.4$\pm$0.3    &    8.4\\  
T313    &0  & 1695    & 1657$\pm$33  &-38    &5.0$\pm$0.3    &    12.8\\  
T324    &0  & 1690    & 1679$\pm$24  &-11    &5.2$\pm$0.3   &      7.7\\
T343    &0  & 1630    & 1613$\pm$16  &-17    &5.4$\pm$0.3    &      8.8\\
T352    &0  & 1640    & 1679$\pm$24  &+39    &5.7$\pm$0.3     &    --\\
T365    &0  & 1630    & 1572$\pm$15  &-58    &5.3$\pm$0.3    &      4.3\\
T367    &0  & 1630    & 1657$\pm$13  &+26    &5.2$\pm$0.3      &    6.6\\
T390    &1  & 1530:   & 1689$\pm$35  &+159:   &5.4$\pm$0.3    &    8.9\\  
T395    &1  & 1580:   & 1727$\pm$42  &+147:   &5.3$\pm$0.3     &     7.5\\  
\enddata
\tablenotetext{a}{\,0=agrees with H{\sc i} measurements, 1=does not agree}
\tablenotetext{b}{\,H{\sc i} velocity taken from Hibbard et al.~(2001).}
\tablenotetext{c}{\,A ':' marks clusters where no H{\sc i} was detected at the position of the cluster.  The given velocity was determined from extrapolation from the nearest measured velocity.}

\label{table:properties2}
\end{deluxetable}

%%%%%%%%%%%%%%%%%%%%%%%%%%%%%%%%%%%%%%%%%%%%%%%

%%%%table2%%%%%
%\rotate{
\begin{deluxetable}{lcccccccccccc}
\def\psn{\phs\phn}
\def\pnn{\phn\phn}
%\tablenum{1}
\tablecolumns{12}
\tablewidth{0pt}
\tablecaption{Cluster properties. (The magnitudes have only been corrected for Galactic foreground extinction)}
\tablehead{
\colhead{ID} & \colhead{A/E\tablenotemark{a}} &  \colhead{$\Delta$RA} &  \colhead{$\Delta$DEC}  & \colhead{F336W} & \colhead{F435W} & \colhead{F550M}& \colhead{F814W}& \colhead{F658N}& \colhead{A$_V$} & \colhead{Z}& \colhead{Log(age)}\\
\colhead{ } & \colhead{ } &   \colhead{(J2000) } & \colhead{(J2000) } & \colhead{(mag)} & \colhead{(mag)} & \colhead{(mag)}& \colhead{(mag)}& \colhead{(mag)} & \colhead{(mag)} & \colhead{(\zo) }& \colhead{(year)}  } 
\startdata
T54   &0   &12h01m52.119s   &-18d52m07.3s       &21.10 &       21.53 &       21.15 &       20.30 &       20.65 &1.0    &0.9$\pm$0.1     &6.9$\pm$0.1         \\  
T111  &0   &12h01m53.379s   &-18d51m39.2s       &20.80 &       21.18 &       21.09 &       20.77 &       20.89 &0.0   &0.9$\pm$0.3     &7.9$\pm$0.1           \\  
T130  &0   &12h01m55.360s   &-18d51m38.9s       &20.33 &       20.82 &       20.72 &       20.37 &       20.43 &0.0    &1.0$\pm$0.1     &8.4$\pm$0.1          \\  
T261  &1   &12h01m53.561s   &-18d53m07.9s       &18.90 &20.17 &       20.29 &       20.14 &       18.76 &0.3    &1.1$\pm$0.2     &$<$6.8              \\
T270  &1   &12h01m53.345s   &-18d53m07.6s      &19.61 &       20.14 &       19.70 &       18.91 &       19.38 &1.7    &1.1$\pm$0.2     &$<$6.8              \\
T296  &0   &12h01m52.624s   &-18d53m33.8s      &19.85 &       20.43 &       20.29 &       19.87 &       19.92 &0.2    &1.0$\pm$0.0     &7.9$\pm$0.1            \\  
T297  &0   &12h02m00.112s   &-18d54m33.3s      &...             &      ...        &     22.22\tablenotemark{b}       &     21.60\tablenotemark{b}        & ... & 1.0    &1.1$\pm$0.1\tablenotemark{c}     &8.5$\pm$0.2\tablenotemark{c}  &  \\  
T299  &0   &12h01m52.480s   &-18d53m20.2s      &19.43 &       20.26 &       20.14 &       19.69 &       19.86 &0.2    &0.9$\pm$0.1     &7.35$\pm$0.07          \\  
T313  &0   &12h01m49.744s   &-18d52m21.9s      & 21.29&       21.88 &       21.80 &       21.35 &       21.59 &0.2    &1.0$\pm$0.1     &7.8$\pm$0.1           \\  
T324  &2   &12h01m52.085s   &-18d52m31.9s       &17.76 &       19.01 &       18.97 &       18.74 &       18.40&0.6    &1.2$\pm$0.2     &6.5-6.8\tablenotemark{d}                               \\
T343  &2   &12h01m50.537s   &-18d52m06.6s       &17.23 &       18.43 &       18.44 &       18.30 &       17.73 &0.4    &1.3$\pm$0.2     &6.5-6.8\tablenotemark{d}                               \\
T352  &1   &12h01m53.022s   &-18d52m10.6s       &16.33 &       17.69 &       17.54 &       17.57 &       17.01 &0.3    &1.3$\pm$0.2     &$<$6.8                      \\
T365  &2   &12h01m54.928s   &-18d52m15.4s       & 17.78 &       19.04 &       18.92 &       18.66 &       18.48   & 0.7 &1.1$\pm$0.2     &6.5-6.8\tablenotemark{d}                  \\
T367  &2   &12h01m54.749s   &-18d52m12.9s       & 16.78 &       18.27 &       18.45 &       18.51 &       17.78 & 0.0   &1.3$\pm$0.2     &6.5-6.8\tablenotemark{d}                 \\
T390  &0   &12h01m51.076s   &-18d51m31.5s       &21.37 &       21.50 &       21.35 &       20.94 &       21.15 &0.0    &1.1$\pm$0.4     &8.3$\pm$0.1           \\  
T395  &0   &12h01m50.681s   &-18d51m26.0s       &21.78 &       21.77 &       21.62 &       21.19 &       21.34 &0.1    &1.1$\pm$0.2     &8.8$\pm$0.1           \\  
%T308-jodida saturacion  &1   &55.13  &-129.52 &1238.30$\pm$17.97   &16.3$\pm$0.1  &15.7$\pm$0.1   &1.5   &0.6$\pm$0.2     &$<$6.8 				&12.96 $\pm$5.2  	\\
%g,i
%T297  &0   &60.40  &210.68      &20.3$\pm$0.1  &20.1$\pm$0.1  &1.0    &1.1$\pm$0.1\tablenotemark{c}     &8.5$\pm$0.2\tablenotemark{c}   &       \\  
%T313  &0   &49.64  &80.20       &21.2$\pm$0.1  &21.2$\pm$0.1  &1.1    &1.0$\pm$0.1      &7.8$\pm$0.1           \\  

\enddata
\tablenotetext{a}{\,0=absorption, 1=emission, 2=emission with WR features}
%\tablenotetext{b}{\,From Base position RA=12:01:00 DEC=-18:51:00 (J2000)}
\tablenotetext{b}{\,Photometry from the WFPC2 images with the F555W and F814W filters.}
\tablenotetext{c}{\,Age and metallicity calculated only using H$\gamma$ and [MgFe].}
\tablenotetext{d}{\,Age calculated using WR features.}

\label{table:colors}
\end{deluxetable}
%}

\clearpage


\begin{thebibliography}{99}

%\bibitem[Adams  \& Myers(2001)]{2001ApJ...553..744A} Adams, F.~C., \& Myers, P.~C.\ 2001, \apj, 553, 744 
\bibitem[Anders \& Fritze-v.~Alvensleben(2003)]{2003A&A...401.1063A} Anders, P., \& Fritze-v.~Alvensleben, U.\ 2003, \aap, 401, 1063 

\bibitem[Anders et al.(2007)]{2007MNRAS.377...91A} Anders, P., Bissantz, 
N., Boysen, L., de Grijs, R., \& Fritze-v.~Alvensleben, U.\ 2007, \mnras, 377, 91 

\bibitem[Ashman 
\& Zepf(1992)]{1992ApJ...384...50A} Ashman, K.~M., \& Zepf, S.~E.\ 1992, \apj, 384, 50 

\bibitem[Barnes(1988)]{1988ApJ...331..699B} Barnes, J.~E.\ 1988, \apj, 331, 699 

\bibitem[Barnes(2004)]{2004MNRAS.350..798B} Barnes, J.~E.\ 2004, \mnras, 350, 798 

%\bibitem[Bastian et al.(2005)]{2005A&A...431..905B} Bastian, N., Gieles, M., Lamers, H.~J.~G.~L.~M., Scheepmaker, R.~A., \& de Grijs, R.\ 2005, \aap, 431, 905 

\bibitem[Bastian et al.(2005)]{2005A&A...435...65B} Bastian, N., Hempel, M., Kissler-Patig, M., Homeier, N.~L., \& Trancho, G.\ 2005, \aap, 435, 65 

\bibitem[Bastian et al.(2006)]{2006A&A...445..471B} Bastian, N., Emsellem, E., Kissler-Patig, M., \& Maraston, C.\ 2006a, \aap, 445, 471 

\bibitem[Bastian et al.(2006)]{2006A&A...448..881B} Bastian, N., Saglia, R.~P., Goudfrooij, P., Kissler-Patig, M., Maraston, C., Schweizer, F., \& Zoccali, M.\ 2006b, \aap, 448, 881 
%\bibitem[Bastian et al.(2006)]{2006A&A...448..881B} Bastian, N., Saglia,  R.~P., Goudfrooij, P., Kissler-Patig, M., Maraston, C., Schweizer, F., \&  Zoccali, M.\ 2006a, A\&A, 448, 881 

%\bibitem[Bastian et al.(2007)]{2007MNRAS.379.1333B} Bastian, N.,  Konstantopoulos, I., Smith, L.~J., Trancho, G., Westmoquette, M.~S., \& Gallagher, J.~S.\ 2007, \mnras, 379, 1333 

\bibitem[Bastian et al.(2008)]{2008MNRAS.389..223B} Bastian, N., Gieles,  M., Goodwin, S.~P., Trancho, G., Smith, L.~J., Konstantopoulos, I., \& Efremov, Y.\ 2008, \mnras, 389, 223 

\bibitem[Bastian(2008)]{2008MNRAS.390..759B} Bastian, N.\ 2008, \mnras, 390, 759 

\bibitem[Baumgardt  \& Makino(2003)]{2003MNRAS.340..227B} Baumgardt, H., \& Makino, J.\ 2003, \mnras, 340, 227 

\bibitem[Baumgardt \& Kroupa(2007)]{2007MNRAS.380.1589B} Baumgardt, H., \& Kroupa, P.\ 2007, \mnras, 380, 1589 

\bibitem[Boutloukos \& Lamers(2003)]{2003MNRAS.338..717B} Boutloukos, S.~G., \& Lamers, H.~J.~G.~L.~M.\ 2003, \mnras, 338, 717 

\bibitem[Cappellari \& Emsellem(2004)]{2004PASP..116..138C} Cappellari, M., \& Emsellem, E.\ 2004, \pasp, 116, 138 

\bibitem[Cardiel et  al.(1998)]{1998A&AS..127..597C} Cardiel, N., Gorgas, J., Cenarro, J., \& Gonzalez, J.~J.\ 1998, \aaps, 127, 597 

\bibitem[Chien et al.(2007)]{2007ApJ...660L.105C} Chien, L.-H., Barnes, 
J.~E., Kewley, L.~J., \& Chambers, K.~C.\ 2007, \apjl, 660, L105 

\bibitem[Cox et al.(2008)]{2008MNRAS.384..386C} Cox, T.~J., Jonsson, P., 
Somerville, R.~S., Primack, J.~R., \& Dekel, A.\ 2008, \mnras, 384, 386 

\bibitem[de Grijs et al.(2003)]{2003MNRAS.343.1285D} de Grijs, R., Anders, 
P., Bastian, N., Lynds, R., Lamers, H.~J.~G.~L.~M., 
\& O'Neil, E.~J.\ 2003, \mnras, 343, 1285 

\bibitem[de Grijs 
\& Goodwin(2008)]{2008MNRAS.383.1000D} de Grijs, R., \& Goodwin, S.~P.\ 2008, \mnras, 383, 1000 


\bibitem[Edmunds \& Pagel(1984)]{1984MNRAS.211..507E} Edmunds, M.~G., \& Pagel, B.~E.~J.\ 1984, \mnras, 211, 507 

\bibitem[Elson et al.(1987)]{1987ApJ...323...54E} Elson, R.~A.~W., Fall, 
S.~M., \& Freeman, K.~C.\ 1987, \apj, 323, 54 

\bibitem[Ercolano et al.(2007)]{2007MNRAS.379..945E} Ercolano, B., Bastian, 
N., \& Stasi{\'n}ska, G.\ 2007, \mnras, 379, 945 

\bibitem[Faber et al.(1985)]{1985ApJS...57..711F} Faber, S.~M., Friel, 
E.~D., Burstein, D., \& Gaskell, C.~M.\ 1985, \apjs, 57, 711 

\bibitem[Fall (2004)]{} Fall, S. M. 2004, in ASP Conf. Ser. 322, Formation and Evolution of Massive Young Star Clusters, ed. H. J. G. L. M. Lamers, A. Nota, \& L. J. Smith (San Francisco: ASP), 399

\bibitem[Fall et al.(2005)]{2005ApJ...631L.133F} Fall, S.~M., Chandar, R., 
\& Whitmore, B.~C.\ 2005, \apjl, 631, L133 (FCW05)

\bibitem[Filippenko(1982)]{1982PASP...94..715F} Filippenko, A.~V.\ 1982, 
\pasp, 94, 715 

\bibitem[Gallagher et al.(2001)]{2001AJ....122..163G} Gallagher, S.~C., 
Charlton, J.~C., Hunsberger, S.~D., Zaritsky, D., \& Whitmore, B.~C.\ 2001, \aj, 122, 163 

\bibitem[Gieles et al.(2005)]{2005A&A...441..949G} Gieles, M., Bastian, N., Lamers, H.~J.~G.~L.~M., \& Mout, J.~N.\ 2005, \aap, 441, 949 

%\bibitem[Gieles et al.(2006)]{2006MNRAS.371..793G} Gieles, M., Portegies 
%Zwart, S.~F., Baumgardt, H., Athanassoula, E., Lamers, H.~J.~G.~L.~M., 
%Sipior, M., \& Leenaarts, J.\ 2006, \mnras, 371, 793 

\bibitem[Gieles et al.(2007)]{2007ApJ...668..268G} Gieles, M., Lamers, 
H.~J.~G.~L.~M., \& Portegies Zwart, S.~F.\ 2007, \apj, 668, 268 

\bibitem[Gieles(2008)]{2008arXiv0801.2676G} Gieles, M.\ 2008  in "Young Massive Star Clusters - Initial Conditions and Environments'', 2008, Astrophysics \& Space Science, eds. E. Perez, R. de Grijs, R. M. Gonzalez Delgado (arXiv:0801.2676) 

%\bibitem[Gieles \& Bastian(2008)]{2008A&A...482..165G} Gieles, M., \& Bastian, N.\ 2008, \aap, 482, 165 

%\bibitem[Gieles(2009)]{2009MNRAS.tmp..277G} Gieles, M.\ 2009, \mnras\, in press (arXiv:0901.0830)

\bibitem[Gieles(2009)]{2009MNRAS.394.2113G} Gieles, M.\ 2009, \mnras, 394, 2113 

\bibitem[Gonz{\'a}lez(1993)]{1993PhDT.......172G} Gonz{\'a}lez, J.~J.\ 
1993, Ph.D.~Thesis

\bibitem[Gonz{\'a}lez Delgado et al.(2005)]{2005MNRAS.357..945G}  Gonz{\'a}lez Delgado, R.~M., Cervi{\~n}o, M., Martins, L.~P., Leitherer,  C., \& Hauschildt, P.~H.\ 2005, \mnras, 357, 945 

\bibitem[Goodwin \& Bastian(2006)]{2006MNRAS.373..752G} Goodwin, S.~P., \& Bastian, N.\ 2006, \mnras, 373, 752 

\bibitem[Goudfrooij et al.(2001)]{2001MNRAS.322..643G} Goudfrooij, P., 
Mack, J., Kissler-Patig, M., Meylan, G., \& Minniti, D.\ 2001, \mnras, 322, 643 

%\bibitem[Hibbard \& Mihos(1995)]{1995AJ....110..140H} Hibbard, J.~E., \& Mihos, J.~C.\ 1995, \aj, 110, 140 

\bibitem[Hibbard et al.(2001)]{2001AJ....122.2969H} Hibbard, J.~E., van der 
Hulst, J.~M., Barnes, J.~E., \& Rich, R.~M.\ 2001, \aj, 122, 2969 

\bibitem[King(1962)]{1962AJ.....67..471K} King, I.\ 1962, \aj, 67, 471 

\bibitem[Knapp et al.(2006)]{2006AJ....131..859K} Knapp, G.~R., et al.\ 
2006, \aj, 131, 859 

\bibitem[Knierman et al.(2003)]{2003AJ....126.1227K} Knierman, K.~A., 
Gallagher, S.~C., Charlton, J.~C., Hunsberger, S.~D., Whitmore, B., Kundu, 
A., Hibbard, J.~E., \& Zaritsky, D.\ 2003, \aj, 126, 1227 

\bibitem[Kobulnicky \& Kewley(2004)]{2004ApJ...617..240K} Kobulnicky, H.~A., \& Kewley, L.~J.\ 2004, \apj, 617, 240 

\bibitem[Konstantopoulos et al.(2008)]{2008ApJ...674..846K} 
Konstantopoulos, I.~S., Bastian, N., Smith, L.~J., Trancho, G., 
Westmoquette, M.~S., \& Gallagher, J.~S., III 2008, \apj, 674, 846 

\bibitem[Konstantopoulos et al.(2009)]{iraklis} 
Konstantopoulos, I.~S., Bastian, N., Smith, L.~J., Trancho, G., 
Westmoquette, M.~S., \& Gallagher, J.~S., III 2009, ApJ, in press (arXiv:0906.2006)

\bibitem[Kroupa(2001)]{2001MNRAS.322..231K} Kroupa, P.\ 2001, \mnras, 322, 
231 

\bibitem[Lada \& Lada(2003)]{2003ARA&A..41...57L} Lada, C.~J., \& Lada, E.~A.\ 2003, \araa, 41, 57 

\bibitem[Lamers et al.(2005)]{2005A&A...429..173L} Lamers, H.~J.~G.~L.~M., Gieles, M., \& Portegies Zwart, S.~F.\ 2005, \aap, 429, 173 

\bibitem[Lamers(2008)]{2008arXiv0804.2148L} Lamers, H.~J.~G.~L.~M.\ 2008 in  "Young Massive Star Clusters - Initial Conditions and Environment", Astrophysics and Space Science, Eds. E. Perez, R. de Grijs and R.M. Gonzalez Delgado (arXiv:0804.2148) 

 \bibitem[Larsen(1999)]{1999A&AS..139..393L} Larsen, S.~S.\ 1999, \aaps, 139, 393 

\bibitem[Larsen(2004)]{2004A&A...416..537L} Larsen, S.~S.\ 2004, \aap, 416, 537 

\bibitem[Larsen et al.(2004)]{2004AJ....128.2295L} Larsen, S.~S., Brodie,  J.~P., \& Hunter, D.~A.\ 2004, \aj, 128, 2295 

\bibitem[Larsen(2009)]{2009A&A...494..539L} Larsen, S.~S.\ 2009, \aap, 494, 539 

\bibitem[Leitherer et al.(1999)]{1999ApJS..123....3L} Leitherer, C., et al.\ 1999, \apjs, 123, 3 

\bibitem[Maraston et  al.(2004)]{2004A&A...416..467M} Maraston, C., Bastian, N., Saglia, R.~P., Kissler-Patig, M., Schweizer, F., \& Goudfrooij, P.\ 2004, \aap, 416, 467 

\bibitem[Mengel et al.(2005)]{2005A&A...443...41M} Mengel, S., Lehnert, M.~D., Thatte, N., \& Genzel, R.\ 2005, \aap, 443, 41 

\bibitem[Mengel et al.(2008)]{2008A&A...489.1091M} Mengel, S., Lehnert, M.~D., Thatte, N.~A., Vacca, W.~D., Whitmore, B., \& Chandar, R.\ 2008, \aap, 489, 1091 

\bibitem[Mihos et al.(1993)]{1993ApJ...418...82M} Mihos, J.~C., Bothun, 
G.~D., \& Richstone, D.~O.\ 1993, \apj, 418, 82 

%\bibitem[Meurer et al.(1995)]{1995AJ....110.2665M} Meurer, G.~R., Heckman, 
%T.~M., Leitherer, C., Kinney, A., Robert, C., \& Garnett, D.~R.\ 1995, \aj, 110, 2665 

\bibitem[Miller et al.(1997)]{1997AJ....114.2381M} Miller, B.~W., Whitmore, 
B.~C., Schweizer, F., \& Fall, S.~M.\ 1997, \aj, 114, 2381 

\bibitem[Read et al.(2006)]{2006MNRAS.366..429R} Read, J.~I., Wilkinson,  M.~I., Evans, N.~W., Gilmore, G., \& Kleyna, J.~T.\ 2006, \mnras, 366, 429 

\bibitem[Salpeter(1955)]{1955ApJ...121..161S} Salpeter, E.~E.\ 1955, \apj, 121, 161 

\bibitem[Savage \& Mathis(1979)]{1979ARA&A..17...73S} Savage, B.~D., \& Mathis, J.~S.\ 1979, \araa, 17, 73 

\bibitem[Saviane et al.(2008)]{2008ApJ...678..179S} Saviane, I., Momany, 
Y., da Costa, G.~S., Rich, R.~M., \& Hibbard, J.~E.\ 2008, \apj, 678, 179 

\bibitem[Scheepmaker et  al.(2007)]{2007A&A...469..925S} Scheepmaker, R.~A., Haas, M.~R., Gieles, M., Bastian, N., Larsen, S.~S., \& Lamers, H.~J.~G.~L.~M.\ 2007, \aap, 469, 925 

%\bibitem[Scheepmaker et  al.(2009)]{} Scheepmaker, R.~A. et al.~2009, \aap (submitted)

\bibitem[Schweizer(1987)]{1987nngp.proc...18S} Schweizer, F.\ 1987, Nearly 
Normal Galaxies.~From the Planck Time to the Present, 18 

\bibitem[Schweizer  \& Seitzer(1998)]{1998AJ....116.2206S} Schweizer, F., \& Seitzer, P.\ 1998, \aj, 116, 2206 

\bibitem[Schweizer et al.(2004)]{2004AJ....128..202S} Schweizer, F., 
Seitzer, P., \& Brodie, J.~P.\ 2004, \aj, 128, 202 

\bibitem[Schweizer et al.(2008)]{2008AJ....136.1482S} Schweizer, F., et 
al.\ 2008, \aj, 136, 1482 

\bibitem[Sidoli et al.(2006)]{2006MNRAS.370..799S} Sidoli, F., Smith, 
L.~J., \& Crowther, P.~A.\ 2006, \mnras, 370, 799 

\bibitem[Smith et al.(2007)]{2007ApJ...667L.145S} Smith, L.~J., et al.\ 
2007, \apjl, 667, L145 

\bibitem[Trager et al.(1998)]{1998ApJS..116....1T} Trager, S.~C., Worthey, 
G., Faber, S.~M., Burstein, D., \& Gonzalez, J.~J.\ 1998, \apjs, 116, 1 

\bibitem[Trancho et al.(2007)]{2007ApJ...658..993T} Trancho, G., Bastian, 
N., Schweizer, F., \& Miller, B.~W.\ 2007a, \apj, 658, 993 (T07a) 

\bibitem[Trancho et al.(2007)]{2007ApJ...664..284T} Trancho, G., Bastian, 
N., Miller, B.~W., \& Schweizer, F.\ 2007b, \apj, 664, 284 (T07b)

%\bibitem[Vesperini et al.(2009)]{2009arXiv0904.3934V} Vesperini, E., 
%McMillan, S.~L.~W., \& Portegies Zwart, S.\ 2009, ApJ, in press (arXiv:0904.3934) 

\bibitem[Vesperini et al.(2009)]{2009ApJ...698..615V} Vesperini, E., 
McMillan, S.~L.~W., \& Portegies Zwart, S.\ 2009, \apj, 698, 615 

\bibitem[Werk et al.(2008)]{2008ApJ...678..888W} Werk, J.~K., Putman, 
M.~E., Meurer, G.~R., Oey, M.~S., Ryan-Weber, E.~V., Kennicutt, R.~C., Jr., 
\& Freeman, K.~C.\ 2008, \apj, 678, 888 

\bibitem[Whitmore 
\& Schweizer(1995)]{1995AJ....109..960W} Whitmore, B.~C., \& Schweizer, F.\ 1995, \aj, 109, 960 

\bibitem[Whitmore et al.(1997)]{1997AJ....114.1797W} Whitmore, B.~C., 
Miller, B.~W., Schweizer, F., \& Fall, S.~M.\ 1997, \aj, 114, 1797 

\bibitem[Whitmore et al.(1999)]{1999AJ....118.1551W} Whitmore, B.~C., 
Zhang, Q., Leitherer, C., Fall, S.~M., Schweizer, F., 
\& Miller, B.~W.\ 1999, \aj, 118, 1551 

\bibitem[Whitmore 
\& Zhang(2002)]{2002AJ....124.1418W} Whitmore, B.~C., \& Zhang, Q.\ 2002, \aj, 124, 1418 

\bibitem[Whitmore et al.(2005)]{2005AJ....130.2104W} Whitmore, B.~C., et 
al.\ 2005, \aj, 130, 2104 

\bibitem[Whitmore et al.(2007)]{2007AJ....133.1067W} Whitmore, B.~C., 
Chandar, R., \& Fall, S.~M.\ 2007, \aj, 133, 1067 (WCF07)

\bibitem[Zepf et al.(1999)]{1999AJ....118..752Z} Zepf, S.~E., Ashman, 
K.~M., English, J., Freeman, K.~C., \& Sharples, R.~M.\ 1999, \aj, 118, 752 

\bibitem[Zhang et al.(2001)]{2001ApJ...561..727Z} Zhang, Q., Fall, S.~M., 
\& Whitmore, B.~C.\ 2001, \apj, 561, 727 

\end{thebibliography}
\end{document}